\documentclass[12pt,a4paper]{article}
\usepackage{amssymb}

\usepackage{graphicx}
\usepackage{amsmath}

\newcounter{resultnum}[section]

\newcounter{conclusionnum}[section]

\newcounter{conditionnum}[section]

\newcounter{conjecturenum}[section]

\newcounter{examplenum}[section]

\newcounter{exercisenum}[section]

\newcounter{lemmanum}[section]

\newcounter{notationnum}[section]

\newcounter{theoremnum}[section]

\newcounter{definitionnum}[section]

\newcounter{corollarynum}[section]

\newcounter{remarknum}[section]

\newcounter{propositionnum}[section]

\newcounter{acknowledgementnum}[section]

\newcounter{algorithmnum}[section]

\newcounter{axiomnum}[section]

\newcounter{casenum}[section]

\newcounter{claimnum}[section]

\newcounter{summarynum}[section]

\newcounter{problemnum}[section]

\begin{document}

\title{Einstein--Cartan Algebroids and\\
Black Holes in Solitonic Backgrounds }
\author{ Sergiu I. Vacaru\thanks{%
vacaru@imaff.cfmac.csic.es } \\
-- \\
{\small Instituto de Matem\'aticas y F{\'\i}sica Fundamental,}\\
{\small Consejo Superior de Investigaciones Cient{\'\i}ficas,}\\
{\small Calle Serrano 123, Madrid 28006, Spain }}
\maketitle

\begin{abstract}
We construct a new class of exact solutions describing spacetimes possessing
Lie algebroid symmetry. They are described by generic off--diagaonal 5D
metrics embedded in bosonic string gravity and possess nontrivial limits to
the Einstein gravity. While we focus on nonholonomic vielbein transforms of
the Schwarzschild metrics to 5D ansatz with solitonic backgrounds, much of
the analysis continues to hold for more general configurations with
nontrivial Lie algebroid structure and nonlinear connections. We carefully
investigate some examples when the anchor structure is related to 3D
solitonic interactions. The approach defines a general geometric method of
constructing exact solutions with various type of symmetries and new
developments and applications of the Lie aglebroid theory.


\vskip0.2cm


\vskip0.1cm

2000 AMS Subject Classification:\

17B99, 53A40, 83C20, 83E99
\end{abstract}

\tableofcontents


\section{ Introduction}

It has been widely investigated the black hole geometry and physics for 2D,
3D, 4D and extra dimensional gravity theories \footnote{%
we shall write 2D for two dimensional, 3D for three dimensional and so on...}%
; see, for instance, Refs. \cite{bh4,bhe}. \ The bulk of such exact
solutions define asymptotically flat spacetimes, or spaces with nontivial
cosmological constant, possessing Killing symmetry and constructed for the
Minkowski, or (anti) De Sitter, backgrounds. It is less known how the black
hole objects can be self--consistently defined for non--constant curvature
backgrounds and what kind of symmetries and physical properties may
characterize such solutions. In a series of recent papers, we studied
certain classes of metrics for black ellipsoids \cite{vjhep,v2,vbel}, black
holes and wormholes in solitonic, spinorial, noncommutative and different
types of nonholonomic backgrounds \cite{vsing1,v2} and locally anisotropic
deformations of the Taub-NUT metrics \cite{vts}. The new class of solutions
possess an explicit nonlinear polarization of constants in the metric and
linear connection coefficients, induced from extra dimension and/or by
specific off--diagonal gravitational interactions, and related to a special
class of nonholonomic frame transforms (vielbeins) with associated nonlinear
connection (in brief, N--connection) structure. Such solutions can be
constructed in string/brane gravity, and in general relativity, and
constrained to define asymptotically flat spacetimes. At least in a finite
region, they are characterized by generalized Lie group structure equations
emphasizing certain type of noncommutative symmetries (investigated in Ref. %
\cite{vncs}) or, for a different class of constructions, by Lie algebroid
symmetries.

In this paper, inspired by the Lie algebroid theory with applications to
mechanics and classical field theory \cite{acsw,weins1,lib,mart,dl1} and
recent approaches to the theory of gauge fields and gravity \cite%
{strobl1,strobl2,dlmddv}, we wont to address essentially the following
purposes:

Given an Einstein--Cartan manifold with the metric and affine connections
satisfying the field equations of string gravity, to construct new classes
of exact solutions describing spacetimes with Lie algebroid structure
(gravitational algebroids). Such solutions will be parametrized by generic
off--diagonal metric ansatz, anholonomic moving frames and generalized
affine connections related to certain three/two dimensional solitonic
gravitational configurations\footnote{%
in brief, we shall write 3D; for four dimensions, 4D, and so on...} with
nontrivial limits to the Schwarzschild black hole metric. We shall analyze
the conditions when the solutions can be constrained to describe
gravitational and matter field interactions in the framework of the Einstein
gravity. There will be presented new motivations for the algebroid theory
and developed a new type of Lie algebroid geometry on nonholonomic manifolds
provided with nonlinear connection structure derived from the gravity theory.%
\footnote{%
A manifold is nonholonomic if it is provided with a nonintegrable
distribution of submanifolds. In this paper, we shall emphasize the
constructions for a special class when the distribution defines the
nonlinear connection structure.}

The paper is organized as follows: In section 2, we outline the basic Lie
algebroid constructions behind the Einstein--Cartan spacetimes provided with
nontrivial nonholonomic vielbeins with associated N--connection structure
and consider the effective field equations with string corrections. Section
3 is devoted to a geometric method of constructing 5D and 4D off--diagonal
solutions for gravitational algebroids. In section 4, there are elaborated
two classes of exact solutions with trivial conformal factors generated by
nonholonomic Lie algebroid deformations of the Schwarzschild solution to
certain type of gravitational solitonic spacetimes. The conclusions are
presented in section 5. The Appendix \ref{app1} outlines the main steps of
constructing solutions on nonholonomic manifolds. Additionally to the
solutions investigated in section 4, we briefly analyze two examples of
metrics with nontrivial conformal factors in Appendix \ref{ap2}.

\section{Riemann--Cartan algebroids}

In this paper, for simplicity, we shall work with real, paracompact and
necessary smooth class manifolds and maps and with locally trivial bundle
spaces.

\subsection{Lie algebroids and N--connections}

The standard definition of a Lie algebroid $\mathcal{A}\doteqdot (E,\left[
\cdot ,\cdot \right] ,\rho )$ is related to a vector bundle $\mathcal{E}%
=(E,\pi ,M),$ with a surjective map $\pi :E\longrightarrow M$ of the total
spaces $E$ to the base manifold $M,$ of respective dimensions $\dim E=n+m$
and $\dim M=n.$ The algebroid structure is stated by the anchor map $\rho :\
E\rightarrow TM$ \ ($TM$ is the tangent bundle to $M$) and a Lie bracket on
the $C^{\infty }(M)$--module of sections of $E,$ denoted $Sec(E),$ such that
\begin{equation*}
\left[ X,fY\right] =f\left[ X,Y\right] +\rho (X)(f)Y
\end{equation*}%
for any $X,Y\in Sec(E)$ and $f\in C^{\infty }(M).$ The anchor also induces a
homomorphism of $C^{\infty }(M)$-modules $\rho :Sec(A)\rightarrow \mathcal{X}%
^{1}(M)$ where $\wedge ^{r}(M)$ and $\mathcal{X}^{r}(M)$ will denote,
respectively, the spaces of differential $r$--forms and $r$--multivector
fields on $M.$\footnote{%
on Lie algebroids geometry and applications, see Refs. \cite%
{acsw,weins1,lib,mart,dl1}}

In order to investigate geometric models of gravity and string theories with
nonholonomic frame (vielbein) structure, one does not work on a vector
bundle $E,$ or a tangent bundle $TM,$ but on a general manifold $\mathbf{V}%
,~ $\ $\ dim\mathbf{V}=n+m,$ which is a (pseudo) Riemannian spacetime, or a
more general one with possible torsion and nonmetricity fields (for the
purposes of this paper, we shall consider it to be a Riemann--Cartan, or
Einstein, manifold; see explicit constructions and references in \cite{vjhep}%
). A Lie algebroid structure can be modelled locally on a spacetime $\mathbf{%
V}$ by considering a Whitney type sum%
\begin{equation}
T\mathbf{V=}h\mathbf{V}\oplus v\mathbf{V}  \label{whit}
\end{equation}%
defining a splitting into certain conventional horizontal (h) and vertical
(v) subspaces. In this case, the anchor is defined as a map $\ \widehat{\rho
}:\ \mathbf{V}\rightarrow h\mathbf{V}$ and the Lie bracket structure is
considered on the spaces of sections $Sec(v\mathbf{V}).$ Roughly speaking,
we consider Riemann--Cartan manifolds admitting a locally fibered structure
induced by the splitting (\ref{whit}) when the Lie algebroid constructions
are usual ones but with formal substitutions $E\rightarrow $ $\ \mathbf{V}$
and $M\rightarrow h\mathbf{V.}$ In general, a such structure is not
integrable which mean that we work on a nonholonomic manifold, see details
in Refs. \cite{v0407495,dlmddv}.

The global decomposition (\ref{whit}) is equivalent to an exact sequence%
\begin{equation*}
0\rightarrow v\mathbf{V}\overset{i}{\rightarrow }T\mathbf{V}\rightarrow T%
\mathbf{V}/v\mathbf{V}\rightarrow 0,
\end{equation*}%
giving a morphism $\mathbf{N}:T\mathbf{V}\rightarrow v\mathbf{V}$ such that $%
\mathbf{N}\circ i$ is the unity in the vertical subbundle $v\mathbf{V}$ (the
kernel $\ker \pi ^{\intercal }\doteqdot v\mathbf{V},$ for $\pi ^{\intercal
}:T\mathbf{V}\rightarrow h\mathbf{V})$ where $i:v\mathbf{V}\rightarrow T%
\mathbf{V}$ is the inclusion mapping. The morphism $\mathbf{N}$ defines a
nonlinear connection (in brief, N--connection) on the spacetime $\mathbf{V.}$
The manifolds provided with N--connection structure are called
N--anholonomic \cite{vjhep}.

Let us state the typical notations for abstract (coordinate) indices of
geometrical objects defined with respect to an arbitrary (or coordinate)
local basis, i. e. with respect to a system of reference. For a local basis
on $\mathbf{V,}$ we write $e_{\alpha }=(e_{i},v_{a}).$ The small Greek
indices $\alpha ,\beta ,\gamma ,...$ are considered to be general ones,
running values $1,2,\ldots ,n+m$ and $i,j,k,...$ and $a,b,c,...$
respectively label the geometrical objects on the base and typical ''fiber''
and run, correspondingly, the values $1,2,...,n$ and $1,2,...,m.$ The dual
base is denoted by $e^{\alpha }=(e^{i},v^{a}).$ The local coordinates of a
point $u\in \mathbf{V}$ are written $\mathbf{u=}(x,u),\ $or $u^{\alpha
}=(x^{i},u^{a}),$ where $u^{a}(u)$ is the $a$-th coordinate with respect to
the basis $(v_{a})$ and $(x^{i})$ are local coordinates on $h\mathbf{V}$
with respect to $e_{i}\mathbf{.}$ By $u^{a}(x),$ one denotes sections of $v%
\mathbf{V}$ over $M.$ We shall use ''boldface'' symbols in order to
emphasize that the objects are defined on spaces provided with N--connection
structure.

A N--connection $\mathbf{N}$ \ is described by its coefficients,%
\begin{equation*}
\mathbf{N}=N_{\ \underline{i}}^{\underline{a}}(u)dx^{\underline{i}}\otimes
\frac{\partial }{\partial u^{\underline{a}}}=N_{\ i}^{b}(u)e^{i}\otimes
v_{b},
\end{equation*}%
where we underlined the indices defined with respect to the local coordinate
basis
\begin{equation*}
e_{\underline{\alpha }}=\partial _{\underline{\alpha }}=\partial /\partial
u^{\underline{\alpha }}=(e_{\underline{i}}=\partial _{\underline{i}%
}=\partial /\partial x^{\underline{i}},v_{\underline{a}}=\partial _{%
\underline{a}}=\partial /\partial u^{\underline{a}})
\end{equation*}%
and its dual
\begin{equation*}
e^{\underline{\alpha }}=du^{\underline{\alpha }}=(e^{\underline{i}}=dx^{%
\underline{i}},e^{\underline{a}}=du^{\underline{a}}).
\end{equation*}%
We can also consider a 'vielbein' transform
\begin{equation}
e_{\alpha }=e_{\alpha }^{\ \underline{\alpha }}(\mathbf{u})e_{\underline{%
\alpha }}\mbox{ and }e^{\alpha }=e_{\ \underline{\beta }}^{\alpha }(\mathbf{u%
})e^{\underline{\alpha }}  \label{vilebtr}
\end{equation}%
given respectively by a nondegenerate matrix $e_{\beta }^{\ \underline{%
\alpha }}(\mathbf{u})$ and its inverse $e_{\ \underline{\beta }}^{\alpha }(%
\mathbf{u}).$ Such matrices parametrize maps from a local coordinate frame
and co--frame, respectively, to any general frame $e_{\alpha }=(e_{i},v_{a})$
and co--frame $e^{\alpha }=(e^{i},v^{a}).$ The class of linear connections
consists, for instance, a case of linear dependence on $u^{\underline{a}},$
i. e. $N_{\underline{i}}^{\underline{a}}(x,u)=\Gamma _{\underline{b}%
\underline{i}}^{\underline{a}}(x)u^{\underline{b}}.$

In local form, the Lie algebroid structure on the manifold $\mathbf{V}$ is
stated by its structure functions $\rho _{a}^{i}(x)$ and $C_{ab}^{f}(x)$
defining the relations
\begin{eqnarray}
\ \rho (v_{a}) &=&\rho _{a}^{i}(x)\ e_{i}=\rho _{a}^{\underline{i}}(x)\
\partial _{\underline{i}},  \label{anch} \\
\lbrack v_{a},v_{b}] &=&C_{ab}^{c}(x)\ v_{c}  \label{liea}
\end{eqnarray}%
and subjected to the structure equations
\begin{equation}
\rho _{a}^{j}\frac{\partial \rho _{b}^{i}}{\partial x^{j}}-\rho _{b}^{j}%
\frac{\partial \rho _{a}^{i}}{\partial x^{j}}=\rho _{c}^{j}C_{ab}^{c}~%
\mbox{\ and\ }\sum\limits_{cyclic(a,b,c)}\left( \rho _{a}^{j}\frac{\partial
C_{bc}^{d}}{\partial x^{j}}+C_{af}^{d}C_{bc}^{f}\right) =0.  \label{lasa}
\end{equation}%
For simplicity, we shall omit underlying of coordinate indices if it will
not result in ambiguities. Such equations are standard ones for the Lie
algebroids but defined on N--anholonomic manifolds. In brief, we call such
spaces to be Lie N--algebroids.

By straightforward computations, we can prove that the Lie algebroid and
N--connection structures prescribe a subclass of preferred local frames
related by subclass of matrix transforms (\ref{vilebtr}) linearly depending
on $N_{\ i}^{a}(x,u),$\bigskip\ with the coefficients

\begin{equation}
\mathbf{e}_{\alpha }^{\ \underline{\alpha }}(u)=\left[
\begin{array}{cc}
e_{i}^{\ \underline{i}}(\mathbf{u}) & N_{i}^{b}(\mathbf{u})e_{b}^{\
\underline{a}}(\mathbf{u}) \\
0 & e_{a}^{\ \underline{a}}(\mathbf{u})%
\end{array}%
\right]  \label{vt1}
\end{equation}%
and
\begin{equation}
\mathbf{e}_{\ \underline{\beta }}^{\beta }(u)=\left[
\begin{array}{cc}
e_{\ \underline{i}}^{i\ }(\mathbf{u}) & -N_{k}^{b}(\mathbf{u})e_{\
\underline{i}}^{k\ }(\mathbf{u}) \\
0 & e_{\ \underline{a}}^{a\ }(\mathbf{u})%
\end{array}%
\right] .  \label{vt2}
\end{equation}%
Such transforms generate N--adapted frames%
\begin{equation}
\mathbf{e}_{\alpha }=(\mathbf{e}_{i}=\frac{\partial }{\partial x^{i}}-N_{\
i}^{b}v_{b},v_{b})  \label{dder}
\end{equation}%
\ and dual coframes
\begin{equation}
\mathbf{e}^{\alpha }=(e^{i},\ \mathbf{v}^{b}=v^{b}+N_{\ i}^{b}dx^{i}),
\label{ddif}
\end{equation}%
for any $v_{b}=e_{b}^{\ \underline{b}}\partial _{\underline{b}}$ satisfying
the condition $v_{c}\mathbf{\rfloor }v^{b}=\delta _{c}^{b}.$ In a particular
case, we can take $v_{b}=\partial _{b}.$

We note that the operators $\mathbf{e}_{\alpha }$ (\ref{dder}) \ and $%
\mathbf{e}^{\alpha }$ (\ref{ddif}) are the so--called ''N--elongated''
partial derivatives and differentials which define a N--adapted differential
calculus on N--anholonomic manifolds. In the structure equations (\ref{lasa}%
), we have $e_{i}\rho _{b}^{i}\rightarrow $ $\partial _{i}\rho _{b}^{i}$
because the structure functions $\rho _{a}^{i}(x)$ and $C_{ab}^{f}(x)$ do
not depend on v--variables $u^{a}.$ For trivial N--connections, we can put $%
N_{\ i}^{a}=0$ and obtain the usual Lie algebroid constructions.

The curvature of a N--connection $\mathbf{\Omega \doteqdot -}N_{v}$ is
defined as the Nijenhuis tensor
\begin{equation*}
N_{v}(\mathbf{X,Y})\doteqdot \lbrack \ vX,\ vY]+\ vv[\mathbf{X,Y}]-\ v[\ vX%
\mathbf{,Y}]-\ v[\mathbf{X,}\ vY]
\end{equation*}%
for any $\mathbf{X,Y\in }\mathcal{X}(\mathbf{V})$ associated to the vertical
projection ''$v"$ defined by this N--connection, i. e.
\begin{equation*}
\mathbf{\Omega =}\frac{1}{2}\Omega _{ij}^{b}\ e^{i}\wedge e^{j}\otimes v_{b}
\end{equation*}%
with the coefficients%
\begin{equation*}
\Omega _{ij}^{a}=e_{[j}N_{\ i]}^{a\,}=e_{j}N_{\ i}^{a\,}-e_{i}N_{\
j}^{a\,}+N_{\ i}^{b\,}v_{b}\left( N_{\ j}^{a\,}\right) -N_{\
j}^{b\,}v_{b}\left( N_{\ i}^{a\,}\right) .
\end{equation*}

The vielbeins (\ref{dder}) satisfy certain nonholonomy (equivalently,
anholonomy) relations
\begin{equation}
\left[ \mathbf{e}_{\alpha },\mathbf{e}_{\beta }\right] =W_{\alpha \beta
}^{\gamma }\mathbf{e}_{\gamma }  \label{anhrel}
\end{equation}%
with nontrivial anholonomy coefficients $W_{jk}^{a}=\Omega _{jk}^{a}(x,u),$ $%
W_{ie}^{b}=v_{e}N_{\ i}^{b}(x,u)$ and $W_{ae}^{b}=C_{ae}^{b}(x)$ reflecting
the fact that the Lie algebroid is N--anholonom\-ic.

We can write down the Lie algebroid and N--connection structures in a
compatible form by introducing the ''N--adapted'' anchor%
\begin{equation}
{}\widehat{\mathbf{\rho }}_{a}^{j}(x,u)\doteqdot \mathbf{e}_{\ \underline{j}%
}^{j}(x,u)\mathbf{e}_{a}^{\ \underline{a}}(x,u)\ \rho _{\underline{a}}^{%
\underline{j}}(x)  \label{bfanch}
\end{equation}%
and ''N--adapted'' (boldfaced) structure functions
\begin{equation}
\mathbf{C}_{ag}^{f}(x,u)=\mathbf{e}_{\ \underline{f}}^{f}(x,u)\mathbf{e}%
_{a}^{\ \underline{a}}(x,u)\mathbf{e}_{g}^{\ \underline{g}}(x,u)\ C_{%
\underline{a}\underline{g}}^{\underline{f}}(x),  \label{bfstrf}
\end{equation}%
respectively, into formulas (\ref{anch}), (\ref{liea}) and (\ref{lasa}). We
conclude that the Lie algebroids on N--anholonomic manifolds are defined by
the corresponding sets of functions ${}\widehat{\mathbf{\rho }}_{a}^{j}(x,u)$
and $\mathbf{C}_{ag}^{f}(x,u)$ with additional dependencies on v--variables $%
u^{b}$ for the N--adapted structure functions. For such generalized Lie
N--algebroids, the structure relations became
\begin{eqnarray}
{}\widehat{\mathbf{\rho }}(v_{b}) &=&{}\widehat{\mathbf{\rho }}%
_{b}^{i}(x,u)\ e_{i},  \label{anch1d} \\
\lbrack v_{d},v_{b}] &=&\mathbf{C}_{db}^{f}(x,u)\ v_{f}  \label{lie1d}
\end{eqnarray}%
and the structure equations of the Lie N--algebroid are written
\begin{eqnarray}
{}\widehat{\mathbf{\rho }}_{a}^{j}e_{j}({}\widehat{\mathbf{\rho }}%
_{b}^{i})-{}\widehat{\mathbf{\rho }}_{b}^{j}e_{j}({}\widehat{\mathbf{\rho }}%
_{a}^{i}) &=&{}\widehat{\mathbf{\rho }}_{e}^{j}\mathbf{C}_{ab}^{e},
\label{lased} \\
\sum\limits_{cyclic(a,b,e)}\left( {}\widehat{\mathbf{\rho }}_{a}^{j}e_{j}(%
\mathbf{C}_{be}^{f})+\mathbf{C}_{ag}^{f}\mathbf{C}_{be}^{g}-\mathbf{C}%
_{b^{\prime }e^{\prime }}^{f^{\prime }}{}\widehat{\mathbf{\rho }}_{a}^{j}%
\mathbf{Q}_{f^{\prime }bej}^{fb^{\prime }e^{\prime }}\right) &=&0,  \notag
\end{eqnarray}%
for $\mathbf{Q}_{f^{\prime }bej}^{fb^{\prime }e^{\prime }}=\mathbf{e}_{\
\underline{b}}^{b^{\prime }}\mathbf{e}_{\ \underline{e}}^{e^{\prime }}%
\mathbf{e}_{f^{\prime }}^{\ \underline{f}}~e_{j}(\mathbf{e}_{b}^{\
\underline{b}}\mathbf{e}_{e}^{\ \underline{e}}\mathbf{e}_{\ \underline{f}%
}^{f})$ with the values $\mathbf{e}_{\ \underline{b}}^{b^{\prime }}$ and $%
\mathbf{e}_{f^{\prime }}^{\ \underline{f}}$ defined by the \ N--connection.
The Lie N--algebroid structure is characterized by the coefficients ${}%
\widehat{\mathbf{\rho }}_{b}^{i}(x,u)$ and $\mathbf{C}_{db}^{f}(x,u)$ stated
with respect to the N--adapted frames (\ref{dder}) \ and (\ref{ddif}).

A Riemann--Cartan algebroid (in brief, RC--algebroid) is a Lie algebroid $\
\mathcal{A}\doteqdot (\mathbf{V},\left[ \cdot ,\cdot \right] ,\rho )$
associated to a N--anholonomic manifold $\mathbf{V}$ provided with a
N--connection $\mathbf{N},$ symmetric metric $\mathbf{g(u)}$ and linear
connection $\mathbf{\Gamma (u)}$ structures resulting in a metric compatible
covariant derivative $\mathbf{D},$ when $\mathbf{Dg=0,}$ but, in general,
with non--vanishing torsion. \footnote{%
We consider that reader is familiar with the main concepts from differential
geometry; in general form, for N--anholonomic spaces, the torsion is defined
bellow by the formula (\ref{tors}).} In this work, we shall investigate some
classes of metrics $\mathbf{g(u)}$ and linear connections $\mathbf{\Gamma (u)%
}$ modelling RC--algebroids as exact solutions of the field equations in
string or Einstein gravity.

\subsection{Metrics and linear connections on RC--algebroids}

Let us consider a metric tensor $\mathbf{g}$ on the manifold $\mathbf{V}$
with the coefficients defined with respect to a local coordinate co--basis $%
du^{\alpha }=\left( dx^{i},du^{a}\right) ,$
\begin{equation*}
\mathbf{g}=\underline{g}_{\alpha \beta }(\mathbf{u})du^{\alpha }\otimes
du^{\beta }
\end{equation*}%
where
\begin{equation}
\underline{g}_{\alpha \beta }=\left[
\begin{array}{cc}
g_{ij}+N_{i}^{a}N_{j}^{b}h_{ab} & N_{j}^{e}h_{ae} \\
N_{i}^{e}h_{be} & h_{ab}%
\end{array}%
\right] .  \label{ansatzc}
\end{equation}%
Performing the vielbein transform $\mathbf{e}_{\alpha }=\mathbf{e}_{\alpha
}^{\ \underline{\alpha }}\partial _{\underline{\alpha }}$ and $\mathbf{e}_{\
}^{\beta }=\mathbf{e}_{\ \underline{\beta }}^{\beta }du^{\underline{\beta }}$
with the matrix coefficients defined respectively by (\ref{vt1}) and (\ref%
{vt2}), we write equivalently the metric $\mathbf{g}$ in the form
\begin{equation}
\mathbf{g}=\mathbf{g}_{\alpha \beta }\left( \mathbf{u}\right) \mathbf{e}%
^{\alpha }\otimes \mathbf{e}^{\beta }=g_{ij}\left( \mathbf{u}\right)
e^{i}\otimes e^{j}+h_{cb}\left( \mathbf{u}\right) \ v^{c}\otimes \ v^{b},
\label{dmetrgr}
\end{equation}%
where $g_{ij}\doteqdot \mathbf{g}\left( e_{i},e_{j}\right) $ and $%
h_{cb}\doteqdot \mathbf{g}\left( v_{c},v_{b}\right) $ and $\mathbf{e}_{\nu
}=(e_{i},v_{b})$ and $\mathbf{e}^{\mu }=(e^{i},v^{b})$\ are, respectively,
just the vielbeins (\ref{dder}) and (\ref{ddif}).

If the manifold $\mathbf{V}$ is (pseudo) Riemannian, there is a unique
linear connection (the Levi--Civita connection) $\nabla ,$ which is metric, $%
\nabla \mathbf{g=0,}$ and torsionless, $\ ^{\nabla }T=0,$ but this
connection is not adapted to the nonintegrable distribution induced by $%
N_{i}^{b}(\mathbf{u}).$ In order to construct exact solutions parametrized
by generic off--diagonal metrics, or to investigate nonholonomic frame
structures in gravity models with nontrivial torsion, it is more convenient
to work with more general classes of linear connections which are N--adapted
but contain nontrivial torsion coefficients because of nontrivial
nonholonomy coefficients $W_{\alpha \beta }^{\gamma }.$ For a corresponding
subset of constraints, the solutions can be related to N--anholonomic
configurations in general relativity, see discussions and references from %
\cite{vjhep,v2}.

A distinguished connection (d--connection) $\mathbf{D}=\{\mathbf{\Gamma }%
_{\beta \gamma }^{\alpha }\}$ on $\mathbf{V}$ is a linear connection
conserving under parallelism the Whitney sum (\ref{whit}). This mean that a
d--connection $\mathbf{D}$ may be represented by h- and \ v--components in
the form $\mathbf{\Gamma }_{\beta \gamma }^{\alpha }=\left(
L_{jk}^{i},L_{bk}^{a},B_{jc}^{i},B_{bc}^{a}\right) ,$ stated with respect to
N--elongated frames (\ref{ddif}) and (\ref{dder}), defining a N--adapted
splitting into h-- and v--covariant derivatives, $\mathbf{D}=hD+vD,$ where $%
hD=(L,L)$ and $vD=(B,B).$

A distinguished tensor (in brief, d--tensor; for instance, a d--metric (\ref%
{dmetrgr})) formalism and d--covariant differential and integral calculus
can be elaborated \cite{v2} for spaces provided with general N--connection,
d--connection and d--metric structure by using the mentioned type of
N--elongated operators. The simplest way to perform a d--tensor covariant
calculus is to use N--adapted differential forms with the coefficients
defined with respect to (\ref{ddif}) and (\ref{dder}), for instance, $%
\mathbf{\Gamma }_{\beta }^{\alpha }=\mathbf{\Gamma }_{\beta \gamma }^{\alpha
}\mathbf{e}^{\gamma }.$

The torsion
\begin{equation}
\mathcal{T}^{\alpha }\doteqdot \mathbf{De}^{\alpha }=d\mathbf{e}^{\alpha
}+\Gamma _{\beta }^{\alpha }\wedge \mathbf{e}^{\beta }  \label{tors}
\end{equation}%
of a d--connection $\mathbf{D}$ has the irreducible h- v-- components
(d--torsions),%
\begin{eqnarray}
T_{\ jk}^{i} &=&L_{\ [jk]}^{i},\ T_{\ ja}^{i}=-T_{\ aj}^{i}=B_{\ ja}^{i},\
T_{\ ji}^{a}=\Omega _{\ ji}^{a},\   \notag \\
T_{\ bi}^{a} &=&T_{\ ib}^{a}=v_{b}(N_{i}^{a})-L_{\ bi}^{a},\ T_{\
bc}^{a}=B_{\ [bc]}^{a}.  \label{dtors}
\end{eqnarray}%
This is the result of a straightforward calculation.

On RC--algebroids, the Levi--Civita linear connection\footnote{%
by definition, satisfying the metricity and zero torsion conditions} $\nabla
=\{^{\nabla }\mathbf{\Gamma }_{\beta \gamma }^{\alpha }\}$ is also not
adapted to the global splitting (\ref{whit}). Such nonholonomic manifolds
can be characterized by a different type of linear connections: For
instance, there is a preferred canonical d--connection structure$\ \widehat{%
\mathbf{\Gamma }}$ constructed only from the metric and N--connection
coefficients $[g_{ij},h_{ab},N_{i}^{a}]$ and satisfying the metricity
conditions $\widehat{\mathbf{D}}\mathbf{g}=0$ and $\widehat{T}_{\ jk}^{i}=0$
and $\widehat{T}_{\ bc}^{a}=0.$ This can be checked by straightforward
calculations with respect to the N--adapted bases (\ref{ddif}) and (\ref%
{dder}) if we take
\begin{equation}
\widehat{\mathbf{\Gamma }}_{\beta \gamma }^{\alpha }=\ ^{\nabla }\mathbf{%
\Gamma }_{\beta \gamma }^{\alpha }+\ \widehat{\mathbf{P}}_{\beta \gamma
}^{\alpha }  \label{cdc}
\end{equation}%
with the deformation d--tensor
\begin{equation*}
\widehat{\mathbf{P}}_{\beta \gamma }^{\alpha
}=(P_{jk}^{i}=0,P_{bk}^{a}=v_{b}(N_{k}^{a}),P_{jc}^{i}=-\frac{1}{2}%
g^{ik}\Omega _{\ kj}^{a}h_{ca},P_{bc}^{a}=0),
\end{equation*}%
where $\ ^{\nabla }\mathbf{\Gamma }_{\beta \gamma }^{\alpha }$ are the
coefficients of the Levi--Civita connection. The torsion of the connection (%
\ref{cdc}) is denoted $\widehat{\mathbf{T}}_{\beta \gamma }^{\alpha }.$ It
should be noted that, in general, the torsion components $\widehat{T}_{\
ja}^{i},\ \widehat{T}_{\ ji}^{a}$ and $\widehat{T}_{\ bi}^{a}$ are not zero.
This is an anholonomic frame (or, equivalently, off--diagonal metric) frame
effect.

In explicit form, the h--v--components of the canonical d--connection $%
\widehat{\mathbf{\Gamma }}_{\ \alpha \beta }^{\gamma }$ $=(\widehat{L}%
_{jk}^{i},\widehat{L}_{bk}^{a},$ $\widehat{B}_{jc}^{i},\widehat{B}%
_{bc}^{a}), $ are given by formulas
\begin{eqnarray}
\widehat{L}_{jk}^{i} &=&\frac{1}{2}g^{ir}\left[
e_{k}(g_{jr})+e_{j}(g_{kr})-e_{r}(g_{jk})\right] ,  \label{candcon} \\
\widehat{L}_{bk}^{a} &=&v_{b}(N_{k}^{a})+\frac{1}{2}h^{ac}\left[
e_{k}(h_{bc})-h_{dc}\ v_{b}(N_{k}^{d})-h_{db}\ v_{c}(N_{k}^{d})\right] ,
\notag \\
\widehat{B}_{jc}^{i} &=&\frac{1}{2}g^{ik}v_{c}(g_{jk}),  \notag \\
\widehat{B}_{bc}^{a} &=&\frac{1}{2}h^{ad}\left[
v_{c}(h_{bd})+v_{b}(h_{cd})-v_{d}(h_{bc})\right] .  \notag
\end{eqnarray}%
They present a 'minimal' generalization of the Levi--Civita connection for
the nonholonomic (pseudo) Riemannian manifolds, which in the case of
nontrivial nonholonomy coefficients $W_{\alpha \beta }^{\gamma }$ (\ref%
{anhrel}), resulting in nontrivial d--torsion components (\ref{dtors}),
consist in a subclass of Riemann--Cartan manifolds provided with N--
connection structure.

The formulas (\ref{anhrel}), (\ref{dtors}) and (\ref{cdc}) and (\ref{candcon}%
) are defined on the nonholonomic spacetime $\mathbf{V}$ and contain the
partial derivative operator $v_{c}=\partial /\partial u^{c}.$ We can
emphasize the Lie N--algebroid structure on the space of sections $Sec(v%
\mathbf{V})$ by substituting $v_{c}=\rho _{c}^{i}(x)\partial /\partial x^{i}
$ (using the anchor map (\ref{anch})), or, in N--adapted form, by working
with ''boldface'' operators $v_{c}\rightarrow \mathbf{v}_{c}=\ \widehat{%
\mathbf{\rho }}_{c}^{i}(x,u)e_{i}$ (see formulas (\ref{anch1d}) and (\ref%
{dder})). A such ''anchoring'' of formulas defines canonical maps for
d--metrics, anholonomic frames, d--connections and d--torsions from $\mathbf{%
V}$ to $Sec(v\mathbf{V}).$ For instance, we can define the anchored map for
the ''contravariant'' h--part of the d--metric (\ref{dmetrgr}),
\begin{equation*}
h^{cb}\left( \mathbf{u}\right) \ v_{c}\otimes \ v_{c}\rightarrow
h^{cb}\left( \mathbf{u}\right) \ \widehat{\mathbf{\rho }}_{c}^{i}\ \ \
\widehat{\mathbf{\rho }}_{b}^{j}\ e_{i}\otimes \ e_{j}
\end{equation*}%
modelling a h--metric $\ ^{N}h^{ij}\doteqdot h^{cb}\left( \mathbf{u}\right)
\ \widehat{\mathbf{\rho }}_{c}^{i}\ \ \widehat{\mathbf{\rho }}_{b}^{j}.$ We
can consider certain canonical anchors $\widehat{\mathbf{\rho }}_{b}^{j}$
when $\ ^{N}h^{ij}=$ $g^{ij}.$

By anchoring the N--elongated differential operators, we can
define and compute (substituting $v_{c}$ by $\
\widehat{\mathbf{\rho }}_{c}^{i}e_{i}$
into (\ref{candcon})) the canonical d--connection $^{\rho }\widehat{\mathbf{%
\Gamma }}_{\ \alpha \beta }^{\gamma }$ on $Sec(v\mathbf{V})$ stating a
canonical map $\widehat{\mathbf{\Gamma }}_{\ \alpha \beta }^{\gamma
}\rightarrow \ ^{\rho }\widehat{\mathbf{\Gamma }}_{\ \alpha \beta }^{\gamma
}.$

\subsection{Curvature on nonholonomic RC--algebroids}

In a similar form, with respect to N--adapted bases, we can compute the h--
v--coefficients of the curvature
\begin{equation*}
\mathbf{R}_{\ \beta }^{\alpha }\doteqdot \mathbf{D\Gamma }_{\beta }^{\alpha
}=d\mathbf{\Gamma }_{\beta }^{\alpha }-\mathbf{\Gamma }_{\beta }^{\gamma
}\wedge \mathbf{\Gamma }_{\gamma }^{\alpha }
\end{equation*}%
(i. e. d--curvatures) of a d--connection $\mathbf{\Gamma }_{\gamma }^{\alpha
},$
\begin{eqnarray}
R_{\ hjk}^{i} &=&e_{k}\left( L_{\ hj}^{i}\right) -e_{j}\left( L_{\
hk}^{i}\right) +L_{\ hj}^{m}L_{\ mk}^{i}-L_{\ hk}^{m}L_{\ mj}^{i}-B_{\
ha}^{i}\Omega _{\ kj}^{a},  \notag \\
R_{\ bjk}^{a} &=&e_{k}\left( L_{\ bj}^{a}\right) -e_{j}\left( L_{\
bk}^{a}\right) +L_{\ bj}^{c}L_{\ ck}^{a}-L_{\ bk}^{c}L_{\ cj}^{a}-B_{\
bc}^{a}\Omega _{\ kj}^{c},  \notag \\
R_{\ jka}^{i} &=&v_{a}\left( L_{\ jk}^{i}\right) -D_{k}\left( B_{\
ja}^{i}\right) +B_{\ jb}^{i}T_{\ ka}^{b},  \label{dcurv} \\
R_{\ bka}^{c} &=&v_{a}\left( L_{\ bk}^{c}\right) -D_{k}\left( B_{\
ba}^{c}\right) +B_{\ bd}^{c}T_{\ ka}^{c},  \notag \\
R_{\ jbc}^{i} &=&v_{c}\left( B_{\ jb}^{i}\right) -v_{b}\left( B_{\
jc}^{i}\right) +B_{\ jb}^{h}B_{\ hc}^{i}-B_{\ jc}^{h}B_{\ hb}^{i},  \notag \\
R_{\ bcd}^{a} &=&v_{d}\left( B_{\ bc}^{a}\right) -v_{c}\left( B_{\
bd}^{a}\right) +B_{\ bc}^{e}B_{\ ed}^{a}-B_{\ bd}^{e}B_{\ ec}^{a}.  \notag
\end{eqnarray}%
The ''anchored'' curvature is computed by the same formulas with $%
v_{c}\rightarrow $ $\ \widehat{\mathbf{\rho }}_{c}^{i}e_{i}$ for any given
d--connection $^{\rho }\mathbf{\Gamma }_{\ \alpha \beta }^{\gamma }.$ For
the curvature of the canonical d--connection, we have to use the anchored
components of (\ref{candcon}), for instance,%
\begin{equation*}
\mathcal{R}_{\ bka}^{c}=\ \ \ \widehat{\mathbf{\rho }}_{a}^{i}e_{i}\left(
^{\rho }L_{\ bk}^{c}\right) -\ ^{\rho }D_{k}\left( ^{\rho }B_{\
ba}^{c}\right) +\ ^{\rho }B_{\ bd}^{c}\ ^{\rho }T_{\ ka}^{c},
\end{equation*}%
where we denote by the ''calligraphic'' symbol $\mathcal{R}$ the
RC--algebroid anchored curvature. In a similar form, we can map all
components of \ (\ref{dcurv}) (we omit such details in this work). This mean
that on a N--anholonomic RC--algebroid $\mathbf{V}$ we can work with the
curvature $\mathbf{R}_{\ \beta }^{\alpha }$ or we can transfer the
constructions on $Sec(v\mathbf{V})$ and work with $\mathcal{R}_{\ \beta
}^{\alpha }.$

Contracting the components of (\ref{dcurv}), we define the Ricci d--tensor
\begin{equation*}
\overleftarrow{\mathbf{R}}_{\alpha \beta }\doteqdot \mathbf{R}_{\ \alpha
\beta \tau }^{\tau }
\end{equation*}%
with h- v--components%
\begin{equation}
R_{ij}\doteqdot R_{\ ijk}^{k},\ \ R_{ia}\doteqdot -R_{\ ika}^{k},\
R_{ai}\doteqdot R_{\ aib}^{b},\ R_{ab}\doteqdot R_{\ abc}^{c},
\label{dricci}
\end{equation}%
and the scalar curvature
\begin{equation*}
\overleftarrow{\mathbf{R}}\doteqdot \mathbf{g}^{\alpha \beta }\overleftarrow{%
\mathbf{R}}_{\alpha \beta }=g^{ij}R_{ij}+h^{ab}R_{ab}.
\end{equation*}%
The Einstein d--tensor is computed in standard form,
\begin{equation*}
\mathbf{E}_{\alpha \beta }=\overleftarrow{\mathbf{R}}_{\alpha \beta }-\frac{1%
}{2}\mathbf{g}_{\alpha \beta }\overleftarrow{\mathbf{R}}.
\end{equation*}%
We shall denote the anchored versions of the Ricci and Einstein d--tensors,
respectively, by $\overleftarrow{\mathcal{R}}_{\alpha \beta }$ and $\mathcal{%
E}_{\alpha \beta }.$

\subsection{String gravity and N--anholonomic manifolds}

Let us consider the strength $\widehat{\mathbf{H}}_{\nu \lambda \rho
}\doteqdot \mathbf{e}_{\nu }\mathbf{B}_{\lambda \rho }+\mathbf{e}_{\rho }%
\mathbf{B}_{\nu \lambda }+\mathbf{e}_{\lambda }\mathbf{B}_{\rho \nu }$
(antysimmetric torsion of the antisymmetric tensor $\mathbf{B}_{\rho \nu }=-%
\mathbf{B}_{\nu \rho }$ from the bosonic model of string theory with dilaton
field $\Phi ,$ see details, for instance, in. \cite{str1,str2}) and
introduce the torsion
\begin{equation*}
\mathbf{H}_{\nu \lambda \rho }\doteqdot \widehat{\mathbf{H}}_{\nu \lambda
\rho }+\widehat{\mathbf{Z}}_{\nu \lambda \rho }
\end{equation*}%
where the deformation%
\begin{equation}
\widehat{\mathbf{Z}}_{\nu \lambda }\doteqdot \widehat{\mathbf{Z}}_{\nu
\lambda \rho }\mathbf{e}^{\rho }=\mathbf{e}_{\lambda }\rfloor \widehat{%
\mathbf{T}}_{\nu }-\mathbf{e}_{\nu }\rfloor \widehat{\mathbf{T}}_{\lambda }+%
\frac{1}{2}(\mathbf{e}_{\nu }\rfloor \mathbf{e}_{\lambda }\rfloor \widehat{%
\mathbf{T}}_{\lambda })\mathbf{e}^{\gamma }  \label{zfun}
\end{equation}%
may be computed by using N--adapted differential forms and the interior
product ''$\rfloor ".$ We denote the energy--momentums of fields by
\begin{equation*}
\widehat{\Upsilon }_{\alpha \beta }=\mathbf{\Sigma }_{\alpha \beta }^{[mat]}+%
\mathbf{\Sigma }_{\alpha \beta }^{[T]}
\end{equation*}%
where$\ \mathbf{\Sigma }_{\alpha \beta }^{[mat]}$ is the source from any
possible matter fields and $\mathbf{\Sigma }_{\alpha \beta }^{[T]}(\widehat{%
\mathbf{T}}_{\nu },\Phi )$ contains contributions of torsion and dilatonic
fields.

The dynamics of sigma model of bosonic string gravity with generic
off--diagonal metrics, effective matter and torsion is defined by the system
of field equations
\begin{eqnarray}
\widehat{\mathbf{R}}_{\alpha \beta }-\frac{1}{2}\mathbf{g}_{\alpha \beta }%
\overleftarrow{\mathbf{\hat{R}}} &=&k\widehat{\Upsilon }_{\alpha \beta },
\label{eecdc1} \\
\ \widehat{\mathbf{D}}^{\nu }(\mathbf{H}_{\nu \lambda \rho }) &=&0,
\label{eecdc1a}
\end{eqnarray}%
where $k=const,$ and the Euler--Lagrange equations for the matter fields are
considered on the background $\mathbf{V.}$

Let us consider a well known ansatz in string theory for the $H$--field when
\begin{equation}
\mathbf{H}_{\nu \lambda \rho }=\widehat{\mathbf{Z}}_{\ \nu \lambda \rho }+%
\widehat{\mathbf{H}}_{\nu \lambda \rho }=\lambda _{\lbrack H]}\sqrt{|\mathbf{%
g}_{\alpha \beta }|}\varepsilon _{\nu \lambda \rho }  \label{ans61}
\end{equation}%
where $\varepsilon _{\nu \lambda \rho }$ is completely antisymmetric and $%
\lambda _{\lbrack H]}=const,$ which satisfies the field equations (\ref%
{eecdc1a}) for $\mathbf{H}_{\nu \lambda \rho }.$ In this work, the ansatz (%
\ref{ans61}) is chosen for a 5D N--anholonomic background with $\widehat{%
\mathbf{Z}}_{\ \nu \lambda \rho }$ defined by the d--torsions for the
canonical d--connection. So, the values $\widehat{\mathbf{H}}_{\nu \lambda
\rho }$ are constrained to solve the equations (\ref{ans61}) for a fixed
value of the cosmological constant $\lambda _{\lbrack H]}$ effectively
modelling corrections from string gravity but, in our case, additionally
deformed by a class of N--anholonomy constrains. As a result, a diagonal
(with respect to (\ref{dder}) and (\ref{ddif})) source $\widehat{\Upsilon }%
_{\alpha \beta }=diag\{\widehat{\Upsilon }_{\alpha }\}$ is parametrized in
the form
\begin{equation*}
\widehat{\Upsilon }_{\alpha \beta }=\{\widehat{\Upsilon }_{1}+\frac{\lambda
_{\lbrack H]}^{2}}{4},\widehat{\Upsilon }_{2}+\frac{\lambda _{\lbrack H]}^{2}%
}{4},\widehat{\Upsilon }_{3}+\frac{\lambda _{\lbrack H]}^{2}}{4},\widehat{%
\Upsilon }_{4}+\frac{\lambda _{\lbrack H]}^{2}}{4},\widehat{\Upsilon }_{5}+%
\frac{\lambda _{\lbrack H]}^{2}}{4}\}
\end{equation*}%
where $\widehat{\Upsilon }_{\alpha }$ are defined by certain matter fields
contributions and $\lambda _{\lbrack H]}^{2}/4$ states string contributions.

In terms of differential forms, the equations (\ref{eecdc1}) are written
\begin{equation}
\eta _{\alpha \beta \gamma }\wedge \widehat{\mathcal{R}}_{\ }^{\beta \gamma
}=\widehat{\Upsilon }_{\alpha },  \label{eecdc2}
\end{equation}%
where, for the volume 4--form $\eta \doteqdot \ast 1$ with the Hodje
operator ''$\ast $'', $\eta _{\alpha }\doteqdot \mathbf{e}_{\alpha }\rfloor
\eta ,$ $\eta _{\alpha \beta }\doteqdot \mathbf{e}_{\beta }\rfloor \eta
_{\alpha },$ $\eta _{\alpha \beta \gamma }\doteqdot \mathbf{e}_{\gamma
}\rfloor \eta _{\alpha \beta },...,\widehat{\mathcal{R}}_{\ }^{\beta \gamma
} $ is the curvature 2--form and $\Upsilon _{\alpha }$ denote all possible
matter sources. The deformation of connection (\ref{cdc}) defines a
deformation of the curvature tensor with respect to the curvature of the
Levi--Civita connection, $\ ^{\nabla }\mathcal{R}_{\ }^{\beta \gamma }.$ The
gravitational field equations (\ref{eecdc2}) transforms into
\begin{equation}
\eta _{\alpha \beta \gamma }\wedge \ ^{\nabla }\mathcal{R}_{\ }^{\beta
\gamma }+\eta _{\alpha \beta \gamma }\wedge \ ^{\nabla }\mathcal{Z}_{\
}^{\beta \gamma }=\widehat{\Upsilon }_{\alpha },  \label{eecdc3}
\end{equation}%
where $^{\nabla }\mathcal{Z}_{\ \ \gamma }^{\beta }=\nabla \mathcal{P}_{\ \
\gamma }^{\beta }+\mathcal{P}_{\ \ \alpha }^{\beta }\wedge \mathcal{P}_{\ \
\gamma }^{\alpha }.$

A subclass of solutions of the gravitational field equations for the
canonical d--connection defines also solutions of the Einstein equations for
the Levi--Civita connection if and only if
\begin{equation}
\eta _{\alpha \beta \gamma }\wedge \ ^{\nabla }\mathcal{Z}_{\ }^{\beta
\gamma }=0  \label{einstconstr}
\end{equation}%
and $\widehat{\Upsilon }_{\alpha }=\ ^{\nabla }\Upsilon _{\alpha },$ (i. e.
the effective source is the same one for both type of connections). Such
constraints have to be imposed in order to select some classes of solutions
in general relativity from certain ones constructed by applying the
canonical d--connection (one has to solve some systems of first order
partial differential equations, or, for certain classes of solutions, the
constraints (\ref{einstconstr}) \ reduce to a system of algebraic equations).

\section{Gravitational Algebroid Configurations}

We introduce a class of generic off--diagonal metrics (depending on 3, or 4,
variables) for which the vacuum Einstein equations with generalizations to a
certain type of string and matter field corrections are completely
integrable. There stated certain parametrizations when such exact solutions
define Lie algebroid configurations.

\subsection{The 5D and 4D ansatz}

We consider a five dimensional (5D) Einstein--Cartan\ spacetime $\mathbf{V}$
provided with a N--connection structure $\mathbf{N}=[N_{i}^{4}(u^{\alpha }),$
$N_{i}^{5}(u^{\alpha })]$ where the local coordinates are labeled $u^{\alpha
}=(x^{i},u^{4}=v,u^{5}),$ for $i=1,2,3.$ Let us formulate some general
conditions when a class of exact solutions of the field equations of the
bosonic string gravity (\ref{eecdc1}) and (\ref{eecdc1a}) depending on
holonomic variables $x^{i}$ and on one anholonomic (equivalently,
anisotropic) variable $u^{4}=v$ can be constructed in explicit form. Every
coordinate from a set $u^{\alpha }$ may be time like, a 3D space one, or
extra dimensional. The partial derivatives will be denoted in the form $%
a^{\times }=\partial a/\partial x^{1},a^{\bullet }=\partial a/\partial
x^{2},a^{\prime }=\partial a/\partial x^{3},a^{\ast }=\partial a/\partial v.$

We consider a class of metrics
\begin{equation}
\mathbf{g}^{[\omega ]}=\omega ^{2}(x^{i},v)\hat{\mathbf{g}}_{\alpha \beta
}\left( x^{i},v\right) du^{\alpha }\otimes du^{\beta },  \label{fmetric}
\end{equation}%
with the coefficients $\hat{\mathbf{g}}_{\alpha \beta }$ parametrized by the
ansatz {\footnotesize
\begin{equation}
\left[
\begin{array}{ccccc}
g_{1}+w_{11}h_{4}+n_{1}^{\ 2}h_{5} & w_{12}h_{4}+n_{1}n_{2}h_{5} &
w_{13}h_{4}+n_{1}n_{3}h_{5} & (w_{1}+\zeta _{1})h_{4} & n_{1}h_{5} \\
w_{21}h_{4}+n_{2}n_{1}h_{5} & g_{2}+w_{22}^{\ }h_{4}+n_{2}^{\ 2}h_{5} &
w_{23}h_{4}+n_{2}n_{3}h_{5} & (w_{2}+\zeta _{2})h_{4} & n_{2}h_{5} \\
w_{31}h_{4}+n_{3}n_{1}h_{5} & w_{32}h_{4}+n_{3}n_{2}h_{5} & g_{3}+w_{33}^{\
}h_{4}+n_{3}^{\ 2}h_{5} & (w_{3}+\zeta _{3})h_{4} & n_{3}h_{5} \\
(w_{1}+\zeta _{1})h_{4} & (w_{2}+\zeta _{2})h_{4} & (w_{3}+\zeta _{3})h_{4}
& h_{4} & 0 \\
n_{1}h_{5} & n_{2}h_{5} & n_{3}h_{5} & 0 & h_{5}%
\end{array}%
\right]  \label{ansatzf}
\end{equation}%
} for $w_{ij}=w_{i}w_{j}+\zeta _{i}\zeta _{j},$ a conformal factor
\begin{equation}
\omega ^{2}(x^{i},v)=\omega _{0}^{2}(x^{2},x^{3})\eta _{0}(x^{i},v),\ \eta
_{0}>0,  \label{coeff1a}
\end{equation}%
and functions
\begin{eqnarray}
g_{i} &=&q_{i}(x^{\widehat{k}})\eta _{i}(x^{\widehat{k}}),\   \label{coeff1b}
\\
&&\mbox{\  for\  }q_{1}=\epsilon =\pm 1,\eta _{1}=1,g_{\widehat{j}}=q_{%
\widehat{j}}(x^{\widehat{k}})\eta _{\widehat{j}}(x^{\widehat{k}});  \notag \\
h_{a} &=&q_{a}(x^{\widehat{k}})\eta _{a}(x^{i},v).  \notag
\end{eqnarray}%
The $N$--coefficients (\ref{dder}) and (\ref{ddif}) are parametrized in the
form $N_{i}^{4}=w_{i}(x^{i},v)$ and $N_{i}^{5}=n_{i}(x^{k},v)$ for the
indices running the values $k,j,...=1,2,3;$ $\widehat{k},\widehat{j},...=2,3$
and $a,b,..=4,5.$ Such 5D metrics possess also a second order anisotropy %
\cite{v2,vjhep} when the $N$--coefficients on the second 'shell' \ (with
four holonomic, $(x^{i},u^{5}),$ and one anholonomic, $u^{4},$ coordinates)
are stated by nontrivial $N_{\hat{{i}}}^{5}=\zeta _{\hat{{i}}}(x^{i},v)$
and, for simplicity, $\zeta _{{5}}=0$ (the indices with 'hat' take values
like $\hat{{i}}=1,2,3,5).$ One shall be considered metrics with smooth
limits $\eta _{0},\eta _{\alpha }\rightarrow 1$ which do not change the
signature. We suppose, that such limits result into certain well known exact
solutions of the Einstein equations (for instance, into the Schwarzschild
metric and/or its imbedding into higher dimensional spacetimes).

The metric (\ref{fmetric}) defined by the ansatz (\ref{ansatzf}) can be
represented equivalently in the form (\ref{dmetrgr}), \
\begin{equation}
\mathbf{g}^{[\omega ]}=\omega ^{2}\left[ \epsilon \left( dx^{1}\right)
^{2}+g_{2}\left( dx^{2}\right) ^{2}+g_{3}\left( dx^{3}\right)
^{2}+h_{4}\left( e^{4}\right) ^{2}+h_{5}\left( e^{5}\right) ^{2}\right] ,
\label{ansatzfd}
\end{equation}%
where%
\begin{equation*}
e^{4}=du^{4}+w_{i}dx^{i}\mbox{ and }e^{5}=du^{5}+n_{i}dx^{i}.
\end{equation*}

The nontrivial components of the 5D Einstein equations defined by the Ricci
d--tensor (\ref{dricci}), $\widehat{\mathbf{R}}_{\alpha \beta }=(\widehat{R}%
_{ij},\widehat{R}_{ia},$ $\widehat{R}_{ai},\widehat{S}_{ab}),$ for the
d--metric (\ref{dmetrgr}) and corresponding canonical d--connection $%
\widehat{\mathbf{\Gamma }}_{\ \alpha \beta }^{\gamma }$ (\ref{candcon}) are
stated by the formulas, see details on a similar calculus in the Appendix to
Ref. \cite{v2},%
\begin{eqnarray}
R_{2}^{2} &=&R_{3}^{3}=\frac{1}{2}R_{1}^{1}=  \label{ep1a} \\
&=&-\frac{1}{2g_{2}g_{3}}[g_{3}^{\bullet \bullet }-\frac{g_{2}^{\bullet
}g_{3}^{\bullet }}{2g_{2}}-\frac{(g_{3}^{\bullet })^{2}}{2g_{3}}%
+g_{2}^{^{\prime \prime }}-\frac{g_{2}^{^{\prime }}g_{3}^{^{\prime }}}{2g_{3}%
}-\frac{(g_{2}^{^{\prime }})^{2}}{2g_{2}}]=-\Upsilon _{4}(x^{2},x^{3}),
\notag \\
S_{4}^{4} &=&S_{5}^{5}=-\frac{1}{2h_{4}h_{5}}\left[ h_{5}^{\ast \ast
}-h_{5}^{\ast }\left( \ln \sqrt{|h_{4}h_{5}|}\right) ^{\ast }\right]
=-\Upsilon _{2}(x^{2},x^{3},v).  \label{ep2a} \\
R_{4i} &=&-w_{i}\frac{\beta }{2h_{5}}-\frac{\alpha _{i}}{2h_{5}}=0,
\label{ep3a} \\
R_{5i} &=&-\frac{h_{5}}{2h_{4}}\left[ n_{i}^{\ast \ast }+\gamma n_{i}^{\ast }%
\right] =0,  \label{ep4a}
\end{eqnarray}%
where
\begin{eqnarray}
\alpha _{i} &=&\partial _{i}{h_{5}^{\ast }}-h_{5}^{\ast }\partial _{i}\ln
\sqrt{|h_{4}h_{5}|},\beta =h_{5}^{\ast \ast }-h_{5}^{\ast }[\ln \sqrt{%
|h_{4}h_{5}|}]^{\ast },  \label{abc} \\
\gamma &=&3h_{5}^{\ast }/2h_{5}-h_{4}^{\ast }/h_{4},  \notag
\end{eqnarray}%
for $h_{4}^{\ast }\neq 0,$ $h_{5}^{\ast }\neq 0\ $(the cases with vanishing $%
h_{4}^{\ast }$ or $h_{5}^{\ast }$ should be analyzed additionally) if there
are satisfied the conditions
\begin{equation}
\hat{{\delta }}_{i}h_{4}=0\mbox{\ and\  }\hat{{\delta }}_{i}\omega =0
\label{conf1}
\end{equation}%
for $\hat{{\delta }}_{i}=\partial _{i}-\left( w_{i}+\zeta _{i}\right)
\partial _{4}+n_{i}\partial _{5}$ when the values $\zeta _{\widetilde{i}%
}=\left( \zeta _{{i}},\zeta _{{5}}=0\right) $ are to be defined for the
solutions of (\ref{conf1}).

Both conditions (\ref{conf1}) are satisfied, for instance, if
\begin{equation}
\omega ^{p_{1}/p_{2}}=h_{4}~(p_{1}\mbox{ and }p_{2}\mbox{ are
integers}),  \label{confq}
\end{equation}%
and $\zeta _{i}$ is the solution of the equations \
\begin{equation}
\partial _{i}\omega -(w_{i}+\zeta _{i})\omega ^{\ast }=0.  \label{confeq}
\end{equation}%
\ Here we note that there are different possibilities to solve the
conditions (\ref{conf1}). For instance, if $\omega =\omega _{1}$ $\omega
_{2},$ we can consider that $h_{4}=\omega _{1}^{p_{1}/p_{2}}$ $\omega
_{2}^{p_{3}/p_{4}}$ $\ $for some integers $p_{1},p_{2},p_{3}$ and $p_{4}.$

The Einstein equations (\ref{eecdc2}) for the ansatz (\ref{ansatzf}) are
compatible for nonvanishing sources and if and only if the nontrivial
components of the source, with respect to the frames (\ref{dder}) and (\ref%
{ddif}), are any functions of type
\begin{equation}
\widehat{\Upsilon }_{2}^{2}=\widehat{\Upsilon }_{3}^{3}=\Upsilon
_{2}(x^{2},x^{3},v),\ \widehat{\Upsilon }_{4}^{4}=\widehat{\Upsilon }%
_{5}^{5}=\Upsilon _{4}(x^{2},x^{3})\text{\mbox{ and }}\widehat{\Upsilon }%
_{1}^{1}=\Upsilon _{2}+\Upsilon _{4}.  \label{emc}
\end{equation}%
This follows from the fact that the nontrivial components of the Einstein
d-tensor $\widehat{\mathbf{G}}_{\ \beta }^{\alpha }$ satisfy the conditions
\begin{equation*}
\widehat{G}_{1}^{1}=-(\widehat{R}_{2}^{2}+\widehat{R}_{4}^{4}),\widehat{G}%
_{2}^{2}=\widehat{G}_{3}^{3}=-\widehat{R}_{4}^{4}(x^{2},x^{3},v),\widehat{G}%
_{4}^{4}=\widehat{G}_{5}^{5}=-\widehat{R}_{2}^{2}(x^{2},x^{3}).
\end{equation*}%
\qquad Parametrizations of sources in the form (\ref{emc}) can be satisfied
for quite general distributions of matter, torsion and dilatonic fields (see
Refs. \cite{v2,vjhep,vts} for an analyzes of such configurations; in this
paper, we shall consider that there are given certain values $\Upsilon _{2}$
and $\Upsilon _{4}$ which vanish in the vacuum cases).

There are proofs \cite{vjhep,v2,vncs}, see also the Appendix, that the
system of gravitational field equations (\ref{eecdc1}) (equivalently, (\ref%
{eecdc2})) for the ansatz (\ref{ansatzf}) and nontrivial components of the
Ricci d--tensor (\ref{dricci}) can be solved in general form if there are
given certain values of functions $g_{2}(x^{2},x^{3})$ (or, inversely, $%
g_{3}(x^{2},x^{3})$), $h_{4}(x^{i},v)$ (or, inversely, $h_{5}(x^{i},v)$) and
of sources $\Upsilon _{2}(x^{2},x^{3},v)$ and $\Upsilon _{4}(x^{2},x^{3}).$

Let us denote the local coordinates $u^{\alpha }=\left( x^{i},u^{a}\right) $
with $x^{i}=\left( x^{1},x^{\widehat{i}}\right) ,$ $x^{\widehat{i}}=\left(
x^{2},x^{3}\right) ,$ $u^{a}=\left( u^{4}=v,u^{5}\right) $ and consider
arbitrary signatures $\epsilon _{\alpha }=\left( \epsilon _{1},\epsilon
_{2},\epsilon _{3},\epsilon _{4},\epsilon _{5}\right) $ (where $\epsilon
_{\alpha }=\pm 1).$ Summarizing the results outlined in the Appendix, for
the nondegenerated cases (when $h_{4}^{\ast }\neq 0$ and $h_{5}^{\ast }\neq
0 $ and, for simplicity, for a trivial conformal factor $\omega ),$ we
formulate an explicit result for 5D exact solutions of the system (\ref{ep1a}%
)--(\ref{ep4a}) and (\ref{confeq}):

Any off--diagonal metric
\begin{eqnarray}
\delta s^{2} &=&\epsilon _{1}(dx^{1})^{2}+\epsilon _{\widehat{k}}g_{\widehat{%
k}}\left( x^{\widehat{i}}\right) (dx^{\widehat{k}})^{2}+  \notag \\
&&\epsilon _{4}h_{0}^{2}(x^{i})\left[ f^{\ast }\left( x^{i},v\right) \right]
^{2}|\varsigma _{\Upsilon }\left( x^{i},v\right) |\left( \delta v\right)
^{2}+\epsilon _{5}f^{2}\left( x^{i},v\right) \left( \delta u^{5}\right) ^{2},
\notag \\
\delta v &=&dv+w_{k}\left( x^{i},v\right) dx^{k},\ \delta
u^{5}=du^{5}+n_{k}\left( x^{i},v\right) dx^{k},  \label{gensol1}
\end{eqnarray}%
with the coefficients being of necessary smooth class, where\ \ $g_{\widehat{%
k}}\left( x^{\widehat{i}}\right) $ is a solution of the 2D equation (\ref%
{ep1a}) for a given source $\Upsilon _{4}\left( x^{\widehat{i}}\right) ,$%
\begin{equation}
\varsigma _{\Upsilon }\left( x^{i},v\right) =\varsigma _{4}\left(
x^{i},v\right) =1-\frac{\epsilon _{4}}{16}h_{0}^{2}(x^{i})\int \Upsilon
_{2}(x^{\widehat{k}},v)[f^{2}\left( x^{i},v\right) ]^{2}dv  \label{aux10}
\end{equation}%
and the N--connection coefficients $N_{i}^{4}=w_{i}(x^{k},v)$ and $%
N_{i}^{5}=n_{i}(x^{k},v),$
\begin{equation}
w_{i}=-\frac{\partial _{i}\varsigma _{\Upsilon }\left( x^{k},v\right) }{%
\varsigma _{\Upsilon }^{\ast }\left( x^{k},v\right) }  \label{gensol1w}
\end{equation}%
and
\begin{equation}
n_{k}=n_{k[1]}\left( x^{i}\right) +n_{k[2]}\left( x^{i}\right) \int \frac{%
\left[ f^{\ast }\left( x^{i},v\right) \right] ^{2}}{\left[ f\left(
x^{i},v\right) \right] ^{2}}\varsigma _{\Upsilon }\left( x^{i},v\right) dv,
\label{gensol1n}
\end{equation}%
define an exact solution of the system of Einstein equations (\ref{ep1a})--(%
\ref{ep4a}) for arbitrary nontrivial functions $f\left( x^{i},v\right) $
(with $f^{\ast }\neq 0),$ $h_{0}^{2}(x^{i})$, $\varsigma _{4[0]}\left(
x^{i}\right) ,$ $n_{k[1]}\left( x^{i}\right) $ and $\ n_{k[2]}\left(
x^{i}\right) ,$ and sources $\Upsilon _{2}(x^{\widehat{k}},v),$ $\Upsilon
_{4}\left( x^{\widehat{i}}\right) $ and any integration constants and
signatures $\epsilon _{\alpha }=\pm 1$ which have to be defined by certain
boundary conditions and physical considerations.

Any metric (\ref{gensol1}) with $h_{4}^{\ast }\neq 0$ and $h_{5}^{\ast }\neq
0$ has the property to be generated by a function of four variables $f\left(
x^{i},v\right) $ with emphasized dependence on the anisotropic coordinate $%
v, $ because $f^{\ast }\doteqdot \partial _{v}f\neq 0$ and by arbitrary
sources $\Upsilon _{2}(x^{\widehat{k}},v),$ $\Upsilon _{4}\left( x^{\widehat{%
i}}\right) .$ The rest of arbitrary functions not depending on $v$ have been
obtained in result of integration of partial differential equations. This
fix a specific class of metrics generated by the relation (\ref{p1}) and the
first formula in (\ref{n}). We can generate also a different class of
solutions with $h_{4}^{\ast }=0$ by considering the second formula in (\ref%
{p2}) and respective formulas in (\ref{n}). The ''degenerated'' cases with $%
h_{4}^{\ast }=0$ but $h_{5}^{\ast }\neq 0$ and inversely, $h_{4}^{\ast }\neq
0$ but $h_{5}^{\ast }=0$ are more special and request a proper explicit
construction of solutions. Nevertheless, such type of solutions are also
generic off--diagonal and they could be of substantial interest.

The sourceless case with vanishing $\Upsilon _{2}$ and $\Upsilon _{4}$ is
defined by the statement: Any off--diagonal metric (\ref{gensol1}) with $%
\varsigma _{\Upsilon }=1,$ $h_{0}^{2}(x^{i})=$ $h_{0}^{2}=const,$ $w_{i}=0$
and $n_{k}$ computed as in (\ref{gensol1n}) but for $\varsigma _{\Upsilon
}=1,$ defines a vacuum solution of 5D Einstein equations for the canonical
d--connection (\ref{candcon}). By imposing additional constraints on
arbitrary functions from $N_{i}^{5}=n_{i}$ and $N_{i}^{5}=w_{i},$ in order
to satisfy the conditions (\ref{einstconstr}), we can select just those
off--diagonal gravitational configurations when the Levi--Civita connection
and the canonical d--connections are related to the same class of solutions
of the vacuum Einstein equations, see details in Ref. \cite{v2}.

Finally, one should be noted that we can reduce the constructions to a 4D
manifold provided with local coordinates $(x^{2},x^{3},u^{4},u^{5})$ if we
exclude dependencies on $x^{1}$ and do not consider terms with indices
taking values $i=1.$ This way, one can be generated exact 4D exact
solutions: for certain parametrizations one get metrics with Lie
N--algebroid symmetries.

\subsection{Gravitational Lie algebroid configurations}

Let us analyze the conditions when a subclass of d--metrics of type (\ref%
{dmetrgr}) (for 5D, a subclass of metrics of type (\ref{ansatzfd})) models a
Lie algebroid provided with N--connection structure.\ We write
\begin{equation*}
\mathbf{g}=\mathbf{g}^{\alpha \beta }\left( \mathbf{u}\right) \mathbf{e}%
_{\alpha }\otimes \mathbf{e}_{\beta }=g^{ij}\left( \mathbf{u}\right)
e_{i}\otimes e_{j}+h^{ab}\left( \mathbf{u}\right) \ v_{b}\otimes \ v_{b},
\end{equation*}%
where $v_{a}\doteqdot $ $e_{a\ }^{\ \underline{a}}(x,u)\ \partial /\partial
u^{\underline{a}}$ satisfy the Lie N--algebroid conditions $%
v_{a}v_{b}-v_{b}v_{a}=\mathbf{C}_{ab}^{d}(x,u)v_{d}$ of type (\ref{lie1d})
with $e_{i}$ being of type (\ref{dder}). Using the anchor map (\ref{anch1d})
with
\begin{equation}
\ \widehat{\mathbf{\rho }}_{a^{\prime }}^{i}\left( \mathbf{u}\right) =%
\mathbf{e}_{\ \underline{i}}^{i}(\mathbf{u})\mathbf{e}_{a^{\prime }}^{\
\underline{a}}(\mathbf{u})\ \rho _{\underline{a}}^{\underline{i}}(x)
\label{anchbf}
\end{equation}%
defined in the form (\ref{bfanch}) by some matrices of type (\ref{vt1}) and (%
\ref{vt2}), we can write the canonical relation
\begin{equation}
g^{ij}\left( \mathbf{u}\right) =h^{a^{\prime }b^{\prime }}\left( \mathbf{u}%
\right) \ \widehat{\mathbf{\rho }}_{a^{\prime }}^{i}\left( \mathbf{u}\right)
\ \widehat{\mathbf{\rho }}_{b^{\prime }}^{j}\left( \mathbf{u}\right) .
\label{cananch1}
\end{equation}%
As a result, the h--component of the d--metric can be represented%
\begin{eqnarray*}
g^{ij}\left( \mathbf{u}\right) e_{i}\otimes e_{j} &=&h^{a^{\prime }b^{\prime
}}\left( \mathbf{u}\right) e_{\ \underline{i}}^{i}\left( \mathbf{u}\right) \
e_{a^{\prime }}^{\ \underline{a}}(x)\rho _{\underline{a}}^{\underline{i}%
}(x)e_{\ \underline{j}}^{j}\left( \mathbf{u}\right) e_{b^{\prime }}^{\
\underline{b}}(x)\rho _{\underline{b}}^{\underline{j}}(x)e_{i}\otimes e_{j}
\\
&=&h^{a^{\prime }b^{\prime }}\left( \mathbf{u}\right) \rho _{a^{\prime }}^{%
\underline{i}}(x)\rho _{b^{\prime }}^{\underline{j}}(x)\frac{\partial }{%
\partial x^{\underline{i}}}\otimes \frac{\partial }{\partial x^{\underline{j}%
}},
\end{eqnarray*}%
where $\rho _{a^{\prime }}^{\underline{i}}(x)=e_{a^{\prime }}^{\ \underline{a%
}}(x)\rho _{\underline{a}}^{\underline{i}}(x).$ We conclude that a metric (%
\ref{dmetrgr}) admits a Lie algebroid type structure with the structure
functions $\rho _{\underline{a}}^{\underline{i}}(x)$ and $C_{\underline{a}%
\underline{b}}^{\underline{d}}(x)$ if and only if the contravariant
h--component of the corresponding d--metric, with respect to the local
coordinate basis, can be parametrized in the form
\begin{equation}
g^{\underline{i}\underline{j}}\left( \mathbf{u}\right) =h^{a^{\prime
}b^{\prime }}\left( \mathbf{u}\right) \rho _{a^{\prime }}^{\underline{i}%
}(x)\rho _{b^{\prime }}^{\underline{j}}(x).  \label{cananch2}
\end{equation}%
The anchor $\rho _{a^{\prime }}^{\underline{i}}(x)$ may be treated as a
vielbein transform depending on $x$--coordinates lifting the horizontal
components of the contravariant metric on the v--subspace. The Lie type
structure functions $C_{ab}^{d}(x,u)$ define certain anholonomy relations
for the basis $v_{a}.$ On a general Lie N--algebroids we shall consider any
set of coefficients $\ \widehat{\mathbf{\rho }}_{a^{\prime }}^{i}\left(
\mathbf{u}\right) $ and $\mathbf{C}_{ab}^{d}(x,u)$ not obligatory subjected
to the data (\ref{anchbf}).

Let us analyze an example of more general conditions when a metric (\ref%
{fmetric}) (equivalently, a d--metric (\ref{ansatzfd})) defines a class of
anchor maps (\ref{cananch1}). On N--anholonomic manifolds, it is more
convenient to work with the N--adapted relations (\ref{cananch1}) than with (%
\ref{cananch2}). \ For effectively diagonal d--metrics, such anchor map
conditions must be satisfied both for $\eta _{\alpha }=1$ and nontrivial
values of $\eta _{\alpha },$ i. e.
\begin{eqnarray}
g^{i} &=&h^{4}\left( \ \widehat{\mathbf{\rho }}_{4}^{i}\right)
^{2}+h^{5}\left( \ \widehat{\mathbf{\rho }}_{5}^{i}\right) ^{2},
\label{eqanch} \\
q^{i} &=&q^{4}\left( \ \widehat{\mathbf{\rho }}_{4}^{i}\right)
^{2}+q^{5}\left( \ \widehat{\mathbf{\rho }}_{5}^{i}\right) ^{2},%
\mbox{\
for\  }\eta _{\alpha }\rightarrow 1,  \notag
\end{eqnarray}%
where $g^{i}=1/g_{i},h^{a}=1/h_{a},\eta ^{\alpha }=1/$ $\eta _{\alpha }$ and
$q^{\alpha }=1/q_{\alpha }.$ The real solutions of (\ref{eqanch}) are
\begin{equation}
\left( \ \widehat{\mathbf{\rho }}_{4}^{i}\right) ^{2}=q^{i}q_{4}H_{4}^{i},\
\left( \ \widehat{\mathbf{\rho }}_{5}^{i}\right) ^{2}=-q^{i}q_{5}H_{4}^{i},
\label{anch5}
\end{equation}%
where%
\begin{equation*}
H_{a}^{i}=\eta _{a}\frac{1-\eta _{4}/\eta _{i}}{\eta _{5}-\eta _{4}},
\end{equation*}%
$\eta _{1}=1,$ for any parametrization (\ref{coeff1b}) with a set of values
of $q^{\alpha }$ and $\eta ^{\alpha }$ for which $\left( \ \widehat{\mathbf{%
\rho }}_{a}^{i}\right) ^{2}>0.$ We note that the conformal factor $\omega
^{2}$ is not related to such equations and their solutions which mean that
the gravitational ''boldfaced'' anchor structures are conformally invariant,
for such classes of metrics. The nontrivial anchor coefficients can be
related to a general solution of type (\ref{gensol1}), (\ref{gensol1w}) and (%
\ref{gensol1n}) by formulas
\begin{equation*}
g_{1}=\epsilon _{1},\ \epsilon _{\widehat{k}}g_{\widehat{k}}(x^{\widehat{i}%
})=q_{\widehat{k}}\eta _{\widehat{k}},\epsilon _{4}h_{0}^{2}[f^{\ast
}]^{2}\left| \zeta _{\Upsilon }\right| =q_{4}\eta _{4},\epsilon
_{5}f^{2}=q_{5}\eta _{5}
\end{equation*}%
with the functions stated in explicit form by considering nonholonomic
deformations of some already known solutions.

The final step in constructing such classes of metrics is to define $\mathbf{%
C}_{ab}^{d}(x,u)$ from the algebraic relations defined by the first equation
in (\ref{lased}) with given values for $\ \widehat{\mathbf{\rho }}_{a}^{i},$
see (\ref{anch5}), and defined N--elongated operators $e_{i}.$ In result,
the second equation in (\ref{lased}) will be satisfied as a consequence of
the first one. This restrict the classes of possible v--frames, $%
v_{b}=e_{b}^{\ \underline{b}}(x,u)\partial /\partial u^{\underline{b}},$
where $e_{b}^{\ \underline{b}}(x,u)$ have to satisfy the algebraic relations
(\ref{lie1d}). We conclude, that the Lie N--algebroid structure impose
certain algebraic constraints on the coefficients of vielbein transforms.
Such spacetimes are with preferred frame structure which can be taken into
account by explicit constructions with respect to N--adapted frames and
distinguishing the anchor $\widehat{\mathbf{\rho }}_{a}^{i}$ and Lie type $%
\mathbf{C}_{ab}^{d}(x,u)$ structures functions.

\section{\ Gravitational Solitonic Schwarzschild Algebroids}

\label{sexsol}We construct in explicit form two classes of solutions of the
gravitational field equations (\ref{ep1a})--(\ref{ep4a}) and (\ref{conf1})
with nontrivial Lie algebroid and N--connection structure describing
generalizations of the 4D Schwarzschild metric to generic off--diagonal 5D
and 4D ansatz with nontrivial 3D gravitational solitonic backgrounds. For
simplicity, we study here d--metrics with trivial conformal factors $\omega
=1$ (see Appendix \ref{ap2} for similar two solutions with $\omega \neq 1).$

As the starting point for our considerations, we consider the 5D metric
\begin{equation}
ds^{2}=\epsilon d\chi ^{2}-a^{2}(p)\left( dp^{2}+d\theta ^{2}+\sin
^{2}\theta d\varphi ^{2}\right) +b^{2}\left( p\right) dt^{2}  \label{schw5}
\end{equation}%
with extra dimension coordinate $\chi $ defining a trivial extension of the
Schwarz\-schild spacetime for
\begin{equation}
a^{2}(p)=a^{2}(\xi )=\frac{\zeta _{g}^{2}}{\xi ^{2}}(\xi
+1)^{2},b^{2}(p)=b^{2}(\xi )=\left( \frac{\xi -1}{\xi +1}\right) ^{2}
\label{schw5coef}
\end{equation}%
with $dp=d\xi /\xi ,$ where $\xi =\zeta /\zeta _{g}$ is related with the
usual radial coordinate by formula $r=\zeta \left( 1+r_{g}/4\zeta \right)
^{2},$ for $\zeta _{g}=r_{g}/4$ with $r_{g}=2G_{[4]}m_{0}/c^{2}$ being the
4D Schwarzschild radius of a point particle of mass $m_{0}$; $%
G_{[4]}=1/M_{P[4]}^{2}$ is the 4D Newton constant expressed via the Planck
mass $M_{P[4]}$ (in general, we may consider that $M_{P[4]}$ may be an
effective 4D mass scale which arises from a more fundamental scale of the
full, higher dimensional spacetime); we set $c=1.$ The 4D part of (\ref%
{schw5}) expressed in terms of functions on $\xi $\ is just the
Schwarzschild solution in \textit{isotropic spherical coordinates} \cite{sw}.

The diagonal metric (\ref{schw5}) defines a vacuum solution of the Einstein
equations for the Levi--Civita connection. A such solution is with Killing
symmetry and asymptotically flat. By N--anholonomic frame transforms (\ref%
{vt1}) and (\ref{vt2}) of this metric, we shall generate new classes of
exact solutions describing Lie N--algebroid configurations with a nontrivial
solitonic background.

\subsection{Stationary off--diagonal solutions}

Let us firstly analyze in details how we can generate a stationary solution
(with the coefficients not depending on the time like variable $u^{5}=t)$ by
deforming nonholonomically the metric (\ref{schw5}) to a generic
off--diagonal metric (\ref{ansatzf}) (equivalently, to a d--metric, (\ref%
{ansatzfd})) with trivial conformal factor $\omega ^{2}=1$ for a set of
local coordinates $(x^{\alpha }=(\chi ,p,\theta ),$ $u^{a}=(v=\varphi ,$ $%
t). $ We write%
\begin{eqnarray}
q_{1} &=&\epsilon \rightarrow g_{1}=\epsilon ,\   \label{data1} \\
q_{2} &=&-a^{2}(p)\rightarrow g_{2}=q_{2}(p)\ \eta _{2}(p,\theta ),\
q_{3}=-a^{2}(p)\rightarrow g_{3}=q_{3}(p)\ \eta _{3}(p,\theta ),  \notag \\
q_{4} &=&-a^{2}(p)\sin ^{2}\theta \rightarrow h_{4}=q_{4}(p,\theta )\eta
_{4}(p,\theta ,\varphi ),  \notag \\
\ q_{5} &=&b^{2}(p)\rightarrow h_{5}=q_{5}(p)\eta _{5}(p,\theta ,\varphi ),
\notag
\end{eqnarray}%
where the non--deformed values are stated by the coefficients (\ref{schw5coef}%
) and the ''polarization'' functions $\eta _{2,3}(p,\theta )$ have to be
found as a solution of type (\ref{auxeq01}), (\ref{auxeq01a}), or (\ref%
{auxeq01b}), depending explicitly of the type of source $\Upsilon
_{4}(p,\theta )$ and vacuum boundary conditions, and the ''polarization''
functions $\eta _{4,5}(p,\theta ,\varphi )$ are solutions of the equations (%
\ref{p1}), or (\ref{p2}), and (\ref{auxf02}), in their turn depending on the
type of source $\Upsilon _{2}(p,\theta ,\varphi )$ and vacuum boundary
conditions. This class of solutions can be represented in the form (\ref%
{gensol1})
\begin{eqnarray}
\delta s^{2} &=&\epsilon (d\chi )^{2}-a^{2}(p)\eta _{2}(p,\theta
)(dp)^{2}-a^{2}(p)\eta _{3}(p,\theta )(d\theta )^{2}  \label{statextsch} \\
&&-h_{0}^{2}(p,\theta )\left[ f^{\ast }\left( p,\theta ,\varphi \right) %
\right] ^{2}|\varsigma _{\Upsilon }\left( p,\theta ,\varphi \right) |\left(
\delta \varphi \right) ^{2}+f^{2}\left( p,\theta ,\varphi \right) \left(
\delta t\right) ^{2},  \notag
\end{eqnarray}%
for
\begin{eqnarray*}
\delta \varphi &=&d\varphi +w_{1}\left( \chi ,p,\theta ,\varphi \right)
d\chi +w_{2}\left( \chi ,p,\theta ,\varphi \right) dp+w_{3}\left( \chi
,p,\theta ,\varphi \right) d\theta ,\  \\
\delta t &=&dt+n_{1}\left( \chi ,p,\theta ,\varphi \right) d\chi
+n_{2}\left( \chi ,p,\theta ,\varphi \right) dp+n_{3}\left( \chi ,p,\theta
,\varphi \right) d\theta ,
\end{eqnarray*}%
where we parametrize
\begin{equation*}
f^{2}\left( p,\theta ,\varphi \right) =b^{2}(p)\eta _{5}(p,\theta ,\varphi
),h_{0}^{2}(p,\theta )=a^{2}(p)/b^{2}(p)
\end{equation*}%
and
\begin{equation*}
h_{0}^{2}(p,\theta )\left[ f^{\ast }\left( p,\theta ,\varphi \right) \right]
^{2}|\varsigma _{\Upsilon }\left( p,\theta ,\varphi \right) |=a^{2}(p)\sin
^{2}\theta \ \eta _{4}(p,\theta ,\varphi )
\end{equation*}%
with the N--connection coefficients $w_{k}$ and $n_{k}$ computed
respectively by the integrals (\ref{gensol1w}) and (\ref{gensol1n}).

If we choose the integration functions to be of sooth class related to
certain distributions of matter, the d--metric (\ref{statextsch}) has the
diagonal coefficients very similar to those for the Schwarzschild metric (%
\ref{schw5}) with the coefficients $a^{2}(p)$ and $b^{2}(p)$ multiplied
respectively on certain polarization $\eta $--functions but (roughly
speaking) embedded into a 5D background of string gravity with nontrivial
torsion and nonholonomic deformation to a preferred frame structure with
associated N--connection. We analyze here an interesting physical case of
non--perturbative gravitational background defined by $f^{2}\left( p,\theta
,\varphi \right) ,$ resulting in a static locally anisotropic polarization $%
\eta _{5}=f^{2}/b^{2}(p),$ related to a soliton solution of the
Kadomtsev--Petviashvili (KdP) equation or (2+1) sine-Gordon (SG) equation
(Refs. \cite{kad,solgr,vjhep} contain original results, basic references and
methods for handling such non--linear equations with solitonic solutions).
In the KdP case, the function $\eta _{5}\left( p,\theta ,\varphi \right) $
satisfies the equation
\begin{equation}
\eta _{5}^{\ast \ast }+\epsilon \left( \dot{\eta}_{5}-6\eta _{5}\eta
_{5}^{\prime }+\eta _{5}^{\prime \prime \prime }\right) ^{\prime }=0,\qquad
\epsilon =\pm 1,  \label{kdp}
\end{equation}%
while in the most general SG case $\eta _{5}(p,\theta ,\varphi )$ satisfies
\begin{equation}
\pm \eta _{5}^{\ast \ast }\mp \ddot{\eta}_{5}\mp \eta _{5}^{\prime \prime
}=\sin (\eta _{5}).  \label{sineq}
\end{equation}%
For simplicity, we can also consider less general versions of the SG
equation where $\eta _{5}$ depends on only one (\textit{e.g.} $\varphi $ and
$x_{1}$) variable. We use the notation $\eta _{5}=\eta _{5}^{stn}$ or $\eta
_{5}=\eta _{5}^{stn}$ with $"stn=KP",$ or $=SG,$ depending if $\eta _{5}$
satisfies equation (\ref{kdp}), or (\ref{sineq}) respectively.

For a stated solitonic form for $h_{5}=h_{5}^{stn}=b^{2}(p)\eta _{5}^{stn},$
with $b^{2}(p)$ taken as for the Schwarzschild metric, $h_{4}$ can be
computed
\begin{equation}
h_{4}=h_{4}^{stn}=h_{[0]}^{2}\left[ \left( \sqrt{|h_{5}^{stn}(p,\theta
,\varphi )|}\right) ^{\ast }\right] ^{2}  \label{p1b}
\end{equation}%
where $h_{[0]}$ is a constant (see formula (\ref{p1}) in the Appendix). This
allows to define $\eta _{4}^{stn}$ $(p,\theta ,\varphi )$ and $%
f^{stn}(p,\theta ,\varphi ),$ which (by using the $f$--function) result in
off--diagonal terms $w_{k}^{stn}\left( \chi ,p,\theta ,\varphi \right) $ (%
\ref{gensol1w}) and $n_{k}^{stn}\left( \chi ,p,\theta ,\varphi \right) $ (%
\ref{gensol1n}). The 3D solitonic character of such N--connection
coefficients can be substantially \ modified by presence of the source $%
\Upsilon _{2}(p,\theta ,\varphi ).$ If $\Upsilon _{2}\rightarrow 0,$ one has
$\varsigma _{\Upsilon }\rightarrow 1$ (\ref{aux10}) and we can put $%
w_{k}^{stn}\rightarrow 0$ but $n_{k}^{stn}$ would preserve solitonic
contributions.

The mentioned 3D solitonic background posses a specific nontrivial torsion
related both to the $B$--field in string gravity with cosmological constant
approximations (\ref{ans61}) and d--torsions (\ref{dtors}) of the canonical
d--connection. We can compute such values by associated to the coefficients
of (\ref{statextsch}) for any type of solitonic (KdP or SG) background. At
the first step one defines the coefficients of the canonical d--connection $%
^{stn}\widehat{\Gamma }_{\beta \gamma }^{\alpha }$ (\ref{candcon}) then of $%
\ ^{stn}\widehat{T}_{\beta \gamma }^{\alpha }$ which allows to compute the
deformations $\ ^{stn}\widehat{\mathbf{Z}}_{\ \nu \lambda \rho }$(\ref{zfun}%
) and, at the second step, we find the N--anholonomically deformed string
torsion $\ $%
\begin{equation*}
^{stn}\widehat{\mathbf{H}}_{\nu \lambda \rho }=\lambda _{\lbrack H]}\sqrt{|\
^{stn}\mathbf{g}_{\alpha \beta }|}\varepsilon _{\nu \lambda \rho }-\ ^{stn}%
\widehat{\mathbf{Z}}_{\ \nu \lambda \rho }
\end{equation*}%
for a stated value of $\lambda _{\lbrack H]}.$ This way, by using the
solitonic coefficients of the d--metric (\ref{statextsch}) and mentioned
procedure of computation of $^{stn}\widehat{\mathbf{H}}_{\nu \lambda \rho }$
we generate a class of exact solutions of the field equations of the bosonic
string gravity (\ref{eecdc1}) \ and (\ref{eecdc1a}) defined by
N--anholonomic maps to 3D soliton backgrounds. As a matter of principle, we
can restrict the polarizations $\eta _{\alpha }$ and the integration
functions of type $h_{0}^{2}(x^{i})$, $\varsigma _{4[0]}\left( x^{i}\right)
, $ $n_{k[1]}\left( x^{i}\right) $ and $\ n_{k[2]}\left( x^{i}\right) $ to
the form satisfying the conditions (\ref{einstconstr}) which in 4D selects
just the Einstein type metrics (vacuum ones or with certain effective matter
modifications). Some new examples and discussion of the former obtained
solutions satisfying such constraints are given in Ref. \cite{v2}. Here we
emphasize that the considered method of N--anholonomic transforms is a
powerful one which states a geometric procedure of constructing general
classes of solutions with generic off--diagonal metrics, torsions and
nonholonomic frames.

We developed our procedure of constructing new solutions in gravity by
starting from the Schwarzschild metric which is asymptotically flat and
posses Killing symmetry. The resulting N--anholonomically deformed metrics
do not have, in general, such properties. As a matter of principle, we can
chose the mentioned set of polarization functions to tend asymptotically
(for a certain effective radial coordinate $r\rightarrow \infty )$ to the
Minkowski 4D metric trivially embedded into 5D, very similarly to the
Schwarzschild case. Nevertheless, even in a such case, our solutions will
have a very different symmetry properties at least in a finite region of the
5D (or 4D) spacetime. In this paper, we investigate the conditions when such
N--anholonomic spacetimes may be characterized by Lie algebroid
configurations. The nontrivial anchor coefficients (\ref{anch5}) are easy to
be computed for such configurations by using \ the data (\ref{data1}) and
any solitonic solution:%
\begin{equation*}
\left( \ \widehat{\mathbf{\rho }}_{4}^{1}\right) ^{2}=-\epsilon a^{2}(p)\sin
^{2}\theta \ H_{5}^{1},\left( \ \widehat{\mathbf{\rho }}_{4}^{2}\right)
^{2}=\sin ^{2}\theta \ H_{5}^{2},\left( \ \widehat{\mathbf{\rho }}%
_{4}^{3}\right) ^{2}=\sin ^{2}\theta \ H_{5}^{3},
\end{equation*}%
\begin{equation*}
\left( \ \widehat{\mathbf{\rho }}_{5}^{1}\right) ^{2}=\epsilon b^{2}(p)\
H_{5}^{1},\left( \ \widehat{\mathbf{\rho }}_{5}^{2}\right) ^{2}=-\frac{%
b^{2}(p)\ }{a_{2}(p)}H_{5}^{2},\left( \ \widehat{\mathbf{\rho }}%
_{5}^{3}\right) ^{2}=-\frac{b^{2}(p)\ }{a_{2}(p)}H_{5}^{3},
\end{equation*}%
where%
\begin{equation*}
H_{a}^{i}(p,\theta ,\varphi )=\eta _{a}^{stn}(p,\theta ,\varphi )\frac{%
1-\eta _{4}^{stn}(p,\theta ,\varphi )/\eta _{i}(p,\theta )}{\eta
_{5}^{stn}(p,\theta ,\varphi )-\eta _{4}^{stn}(p,\theta ,\varphi )},
\end{equation*}%
for $\eta _{1}(p,\theta )=1;a=3,4$ and $i=1,2,3.$ Such anchor coefficients
can be zero, when $b^{2}(p)=0$ and for certain polarizations one could be $%
\left( \ \widehat{\mathbf{\rho }}_{a}^{i}\right) ^{2}<0$ in some spacetime
regions. We have to exclude such regions for the real valued solutions, or
to redefine the type of anchor maps in order to obtain to generate only real
metrics and connections. A Lie algebroid \ configuration is completely
established after fixing a frame of reference (in general, nonholonomic) in
the v--subspace, defining the values $C_{bc}^{a}(p,\theta )$, see (\ref%
{lie1d})).

Finally, in this section, we address this important physical question: The
Schwarzschild solution defines a 4D black hole. Should the N--anholonomic
deformations define similar objects? \ In general,  the black hole character
of the solutions is not preserved under such transforms (not preserving the
spherical Killing symmetry). The singularities of any exact solution can be
investigated by an explicit computation of the components of the Riemann
d--tensors (\ref{dcurv}) with a corresponding anchoring of formulas. We omit
such cumbersome formulas and their analysis in this paper. Nevertheless, it
is almost obvious that there is a subclass of N--anholonomic and/or
solitonic transforms preserving the type of certain black hole
configurations (even the solutions are deformed to generic off--diagonal
configurations). This follows from the fact that in the vicinity of the
black hole singularity we can define some infinitesimal nonholonomic maps
with smooth coefficients which preserve all singular properties of the
curvature but induce certain additional smooth off--diagonal corrections and
the same structure of the metric coefficients but defined with respect to
N--adapted frames. This way we defined the so--called black ellipsoid
solutions \cite{vbel}. For some special cases (see details in Refs. \cite%
{vjhep}), one can select certain black hole configurations on nontrivial
backgrounds, in our case, of stationary solitonic character.

\subsection{Time--depending solitonic backgrounds}

We also deform nonholonomically the metric (\ref{schw5}) to a generic
off--diagonal metric (\ref{ansatzf}) (equivalently, to a d--metric, (\ref%
{ansatzfd})) with trivial conformal factor $\omega ^{2}=1$ but for a set of
local coordinates $x^{\alpha }=(\chi ,p,\theta )$ and $u^{a}=(v=t,$ $\varphi
)$ when
\begin{eqnarray*}
q_{1} &=&\epsilon \rightarrow g_{1}=\epsilon ,\  \\
q_{2} &=&-a^{2}(p)\rightarrow g_{2}=q_{2}(p)\ \eta _{2}(p,\theta ),\
q_{3}=-a^{2}(p)\rightarrow g_{3}=q_{3}(p)\ \eta _{3}(p,\theta ), \\
\ q_{4} &=&b^{2}(p)\rightarrow h_{4}=q_{5}(p)\eta _{4}(p,\theta ,t), \\
q_{5} &=&-a^{2}(p)\sin ^{2}\theta \rightarrow h_{5}=q_{5}(p,\theta )\eta
_{5}(p,\theta ,t).
\end{eqnarray*}%
These data are different from (\ref{data1}) by emphasizing the
''anisotropic'' dependence on the time coordinate $t$ instead of the angular
one, as it was on $\varphi $ for the previous solution. In this case, the
''polarization'' functions $\eta _{4,5}(p,\theta ,t)$ are solutions of the
equations (\ref{p1}), or (\ref{p2}), and (\ref{auxf02}), in their turn
depending on the type of a variable in time source $\Upsilon _{2}(p,\theta
,t)$ and vacuum boundary conditions. This is another class of d--metrics (%
\ref{gensol1}),
\begin{eqnarray}
\delta s^{2} &=&\epsilon (d\chi )^{2}-a^{2}(p)\eta _{2}(p,\theta
)(dp)^{2}-a^{2}(p)\eta _{3}(p,\theta )(d\theta )^{2}  \label{sol2} \\
&&+h_{0}^{2}(p,\theta )\left[ f^{\ast }\left( p,\theta ,t\right) \right]
^{2}|\varsigma _{\Upsilon }\left( p,\theta ,t\right) |\left( \delta t\right)
^{2}-f^{2}\left( p,\theta ,t\right) \left( \delta \varphi \right) ^{2},
\notag
\end{eqnarray}%
for
\begin{eqnarray*}
\delta t &=&dt+w_{1}\left( \chi ,p,\theta ,t\right) d\chi +w_{2}\left( \chi
,p,\theta ,t\right) dp+w_{3}\left( \chi ,p,\theta ,t\right) d\theta , \\
\delta \varphi &=&d\varphi +n_{1}\left( \chi ,p,\theta ,t\right) d\chi
+n_{2}\left( \chi ,p,\theta ,t\right) dp+n_{3}\left( \chi ,p,\theta
,t\right) d\theta ,
\end{eqnarray*}%
where we parametrize
\begin{equation*}
f^{2}\left( p,\theta ,t\right) =b^{2}(p)\eta _{5}(p,\theta
,t),h_{0}^{2}(p,\theta )=b^{2}(p)/a^{2}(p)\sin ^{2}\theta
\end{equation*}%
and
\begin{equation*}
h_{0}^{2}(p,\theta )\left[ f^{\ast }\left( p,\theta ,t\right) \right]
^{2}|\varsigma _{\Upsilon }\left( p,\theta ,t\right) |=b^{2}(p)\ \eta
_{4}(p,\theta ,t)
\end{equation*}%
with the N--connection coefficients $w_{k}$ and $n_{k}$ computed
respectively by the integrals (\ref{gensol1w}) and (\ref{gensol1n}).

The solitonic background of the d--metric (\ref{sol2}) is given by
\begin{equation*}
h_{5}=h_{5}^{stn}=-a^{2}(p)\sin ^{2}\theta \ \eta _{5}^{stn}
\end{equation*}
and
\begin{equation*}
h_{4}=h_{4}^{stn}=h_{[0]}^{2}\left[ \left( \sqrt{|h_{5}^{stn}(p,\theta ,t)|}%
\right) ^{\ast }\right] ^{2}
\end{equation*}%
defined by the 3D solitonic equation (\ref{kdp}), or (\ref{sineq}), for the
new set of coordinates, when $h_{[0]}=const,$ see formula (\ref{p1}) in the
Appendix. This allows to define $\eta _{4}^{stn}$ $(p,\theta ,t)$ and $%
f^{stn}(p,\theta ,t),$ which (by using the $f$--function) result in
off--diagonal terms (i. e. in N--connection coefficients) $w_{k}^{stn}\left(
\chi ,p,\theta ,t\right) $ (\ref{gensol1w}) and $n_{k}^{stn}\left( \chi
,p,\theta ,t\right) $ (\ref{gensol1n}).

This class of metrics is characterized by dynamical anchor maps,%
\begin{equation*}
\left( \ \widehat{\mathbf{\rho }}_{4}^{1}\right) ^{2}=\epsilon b^{2}(p)\
H_{5}^{1},\left( \ \widehat{\mathbf{\rho }}_{4}^{2}\right) ^{2}=-\frac{%
b^{2}(p)\ }{a_{2}(p)}\ H_{5}^{2},\left( \ \widehat{\mathbf{\rho }}%
_{4}^{3}\right) ^{2}=-\frac{b^{2}(p)\ }{a_{2}(p)}\ H_{5}^{3},
\end{equation*}%
\begin{equation*}
\left( \ \widehat{\mathbf{\rho }}_{5}^{1}\right) ^{2}=-\epsilon a^{2}(p)\
\sin ^{2}\theta \ H_{5}^{1},\left( \ \widehat{\mathbf{\rho }}_{5}^{2}\right)
^{2}=\sin ^{2}\theta \ H_{5}^{2},\left( \ \widehat{\mathbf{\rho }}%
_{5}^{3}\right) ^{2}=\sin ^{2}\theta \ H_{5}^{3},
\end{equation*}%
where%
\begin{equation*}
H_{a}^{i}(p,\theta ,t)=\eta _{a}^{stn}(p,\theta ,t)\frac{1-\eta
_{4}^{stn}(p,\theta ,t)/\eta _{i}(p,\theta )}{\eta _{5}^{stn}(p,\theta
,t)-\eta _{4}^{stn}(p,\theta ,t)},
\end{equation*}%
for $\eta _{1}(p,\theta )=1;a=3,4$ and $i=1,2,3.$ We have to exclude some
spacetime regions from consideration (where the solutions became complex
valued) or to redefine the type of anchor maps in order to generate only
real valued metrics. A dynamical Lie algebroid \ configuration is completely
established after fixing a frame of reference (in general, nonholonomic) in
the v--subspace, defining the values $C_{bc}^{a}(p,\theta ),$ see (\ref%
{lie1d}), in a form compatible with (\ref{lased}).

The constructed in this section exact solutions, in general form, depend on
certain type functions on variables $x^{i}$ obtained from the procedure of
integrating systems of partial equations. In the particular case of
Schwarschild solution, the result of such integration were certain constants
which are defined from the boundary conditions like the asymptotic limit to
the Newton potential for the gravitational field, with Killing spherical
symmetry, and asyptotically Minkowski spacetime. A such result with
integration constants is possible for the corresponding diagonal ansatz for
metric reducing the vacuum Einstein equations to an effective nonlinear
second order partial differential equation.

The generic off--diagonal ansatz considered in paper results in systems of
nonlinear partial differential equations. We proved that by corresponding
geometric methods such equations can be solved in a quite general form which
depends not only on arbitrary constants but also on certain classes of
functions depending on one, two,three and four variables (for 5D
configurations). Roughly speaking, this means that we can extend the
Schwarschild solution to a very general background which can be constrained
in various forms in order to describe different type of nonlinear
interactions, for instance, 3D solitons, or certain nonholonomic Lie
algebroid configurations. Nevertheless, even in such cases, the solutions
depends on some functions on $x^{i}.$ This is characteristic for various
classes of solutions of the systems of nonlinear equations. For instance, in
Refs. \cite{strobl1,strobl2}, there are investigated the conditions when a
''resonable'' theory of gravity is defined from a 2D Poisson setting and
related Lie algebroids. In this work, we considered the problem of selecting
''reasonable'' off--diagonal spacetime possessing Lie N--algebroid
symmetries. Such metrics still depend on some classes of functions (we call
them gravitational polarizations). One can fix an explicit system of
reference and choose the Lie algebroid structure functions to have a limit
to certain Lie type structure constants, following some physical
prescriptions on symmetry and boundary conditions, i. e. a particular exact
solution distinguished from a set of general ones. The priority of the
method developed in this work is that the gravitational algebroids can be
defined as certain general nonlinear configurations but not only as
particular ones with integration constants.

For some very special classes of functions $\eta _{\alpha }$, $w_{i}$ and $%
n_{i},$ the d--metric (\ref{sol2}) may define certain black hole like
configurations with polarized and variable in time constants if we constrain
the configurations to mimic the Schwarzschild metric embedded into
solitonically perturbed 5D spacetime. But, in general, such solutions do not
possess a black hole character and describe a nonlinear gravitational
solitonic dynamics related to a Lie N--algebroid configuration. In the
Appendix \ref{ap2} there are analyzed two classes of d--metrics (\ref%
{ansatzfd}) with nontrivial conformal factor $\omega .$

\section{Conclusions and Discussion}

In this paper, we have examined a new class of 5D metric ansatz defining
exact solutions in string gravity and possessing nontrivial 4D vacuum and
nonvacuum limits to the Einstein gravity. Such models define spacetimes
characterized by Lie algebroid symmetries and prescribed vielbein structures
with associated nonlinear connections. While our analyzis has mainly focused
on the properties of nonholonomic deformations of the Schwarazschild metric
to generic off--diagonal solutions with 3D solitonic configurations, much of
the constructions hold true for more general background metrics and
symmetries. We parametrized such solutions in general form possessing an
explicit dependence on arbitrary integration functions (on 1,2, or 3
variables) and constants. We computed the nontrivial coefficients of the
nonlinear connections and anchor maps defining nonholonomic Lie algebroid
structures. Stating the systems of reference and the boundary conditions, we
can define in explicit form the polarization of constants and metric
coefficients induced by extra dimensional and/or generic off--diagonal
solitonic gravitational interactions and nonholonomic constraints.

The bulk of astrophysical and cosmological applications of exact solutions
are related to spherically symmetric and asymptotically Minkowski
spacetimes. Such constructions are technically more easy to be handled and
generalized to extra dimensions and/or quantum gravity. In another turn,
nonlinear gravitational and matter field configurations and interactions,
with string/brane corrections depending on 2--4 variables and with
generalized (non--Killing) symmetries are very important for elaboration new
types of models with non--perturbative vacuum in modern high energy physics
and gravity. In this work we emphasized possible 3D solitonic deformations
of the Schwarzschild solutions in the presence of extra dimensions,
off--diagonal/ non\-holonomic interactions and bosonic string corrections.

One should worry that arbitrary nonholonomic deformations to certain Lie
algebroid spacetimes will not preserve the former black hole character of
the solution. Nevertheless, it is always possible to define a smooth
subclass of such solitonic deformations which preserve the singular
structure of the curvature tensor but 'slightly' modify the horizons,
polarize the constants and move the black hole solutions on the extra
dimension and/or time like coordinates.

The mentioned approach requests a more sophisticate geometric techniques and
methods. The geometry of nonlinear connections and moving frames have in
this case a new realization in terms of Lie algebroid structures related to
the symmetries of gravitational field equations. This suggests new
directions of investigation both in Lie algebroid mathematics and the
geometry of classical and quantum fields.

We briefly comment and compare our results with the previous applications of
algebroid methods in mechanics and classical field theory \cite%
{weins1,lib,mart,dl1}, and in string, gravity and gauge theories \cite%
{strobl1,strobl2,dlmddv}. It should be emphasized that our gravitational
algebroid constructions are derived as exact solutions from string and
Einstein gravity being elaborated for nonholonomic manifolds \cite%
{vjhep,v2,v0407495}. They are quite different from the usual Lie algebroids
defined for vector or (co) tangent bundles with trivial (vanishing)
nonlinear connection structure \cite{acsw}. \ The first applications of Lie
algebroids to mechanics and jet extensions of Lie algebroids for classical
fields were performed by using the fiber bundle formalism and geometrization
of the Euler--Lagrange equations (for instance, in terms of Poincare--Cartan
forms, or Ehressman connections). In our investigations, we emphasized that
any Lagrangian/Hamiltonian configuration results in canonical nonlinear
connection and adapted metric and linear connection structures which
transform the Lie algebroid constructions to be nonholonomic ones. In this
paper, and in Refs. \cite{v0407495,dlmddv}, we gave certain explicit
examples when the commutative and noncommutative geometric configurations
and almost sympletic and algebroid structures can be derived as exact
solutions in gravity. This presents additional arguments and a knew
understanding of algebroid geometry and certain applications for
constructing new kinds of gauge theories by replacing Lie algebras by Lie or
Courant algebroids and searching for a reasonable theory of gravity.

The final conclusion of this paper is that one could be constructed certain
Lie algebroid models of spacetimes defined as exact solutions in gravity but
such generic off--diagonal gravitational configurations are generated as
nonholonomic manifolds, i. e. as spacetimes provided with nonintegrable
distributions possessing new types of symmetries and nonlinear connection
structure. To investigate the geometric and physical properties of such
vacuum and nonvacuum gravitational algebroid spacetimes is one of the aims
of our further researches.

{\vskip 4pt}

\textbf{Acknowledgement: } The work is supported by a sabbatical fellowship
of the Ministry of Education and Research of Spain.

\appendix

\section{N--anholonomic Deformations and Exact Solutions}

\label{app1}In a series of papers, see \cite{vjhep,v2} and presented there
references, we elaborated the anholonomic frame method of constructing exact
solutions with generic off--diagonal metrics (depending on 2-4 variables) in
general relativity, gauge gravity and certain extra dimension
generalizations. In this Appendix, we outline the results which are
necessary for constructing exact solutions characterized by Lie algebroid
symmetries and nontrivial nonlinear connection (N--connection) structure.

We sketch five steps of generating solutions of the system of second order
nonlinear partial differential equations (\ref{ep1a})--(\ref{ep4a}) and (\ref%
{confeq}) for given certain values of functions $g_{2}(x^{2},x^{3})$ (or,
inversely, $g_{3}(x^{2},x^{3})),\ h_{4}\left( x^{i},v\right) $ (or,
inversely, $h_{5}\left( x^{i},v\right) ),$ $\omega \left( x^{i},v\right) $
and of sources $\Upsilon _{2}(x^{2},x^{3},v)$ and $\Upsilon
_{4}(x^{2},x^{3}):$

\begin{enumerate}
\item The general solution of equation (\ref{ep1a}) may be represented in
the form
\begin{equation}
\varpi =g_{[0]}\exp [a_{2}\widetilde{x}^{2}\left( x^{2},x^{3}\right) +a_{3}%
\widetilde{x}^{3}\left( x^{2},x^{3}\right) ],  \label{solricci1a}
\end{equation}%
were $g_{[0]},a_{2}$ and $a_{3}$ are some constants and the functions $%
\widetilde{x}^{2,3}\left( x^{2},x^{3}\right) $ define any coordinate
transforms $x^{2,3}\rightarrow \widetilde{x}^{2,3}$ for which the 2D line
element becomes conformally flat, i. e.
\begin{equation}
g_{2}(x^{2},x^{3})(dx^{2})^{2}+g_{3}(x^{2},x^{3})(dx^{3})^{2}\rightarrow
\varpi (x^{2},x^{3})\left[ (d\widetilde{x}^{2})^{2}+\epsilon (d\widetilde{x}%
^{3})^{2}\right] ,  \label{con10}
\end{equation}%
where $\epsilon =\pm 1$ for a corresponding signature. For the coordinates $%
\widetilde{x}^{2,3},$ the equation (\ref{ep1a}) transform into%
\begin{equation*}
\varpi \left( \ddot{\varpi}+\varpi ^{\prime \prime }\right) -\dot{\varpi}%
-\varpi ^{\prime }=2\varpi ^{2}\Upsilon _{4}(\tilde{x}^{2},\tilde{x}^{3})
\end{equation*}%
or%
\begin{equation}
\ddot{\psi}+\psi ^{\prime \prime }=2\Upsilon _{4}(\tilde{x}^{2},\tilde{x}%
^{3}),  \label{auxeq01}
\end{equation}%
for $\psi =\ln |\varpi |.$ The form of solutions of (\ref{auxeq01}) depends
on the source $\Upsilon _{4}.$ As a particular case we can consider that $%
\Upsilon _{4}=0.$ We also can prescribe that $g_{2}=g_{3}$ and get the
equation (\ref{auxeq01}) for $\psi =\ln |g_{2}|=\ln |g_{3}|.$ If we select
the case when $g_{2}^{^{\prime }}=0,$ for a given $g_{2}(x^{2}),$ we obtain
from (\ref{ep1a})%
\begin{equation}
g_{3}^{\bullet \bullet }-\frac{g_{2}^{\bullet }g_{3}^{\bullet }}{2g_{2}}-%
\frac{(g_{3}^{\bullet })^{2}}{2g_{3}}=2g_{2}g_{3}\Upsilon _{4}(x^{2},x^{3})
\label{auxeq01a}
\end{equation}%
which can be integrated explicitly for given values of $\Upsilon _{4}.$
Similarly, we can generate solutions for a prescribed $g_{3}(x^{3})$ in the
equation
\begin{equation}
g_{2}^{^{\prime \prime }}-\frac{g_{2}^{^{\prime }}g_{3}^{^{\prime }}}{2g_{3}}%
-\frac{(g_{2}^{^{\prime }})^{2}}{2g_{2}}=2g_{2}g_{3}\Upsilon
_{4}(x^{2},x^{3}).  \label{auxeq01b}
\end{equation}%
We note that a transform (\ref{con10}) is always possible for 2D metrics and
the explicit form of solutions depends on chosen system of 2D coordinates
and on the signature $\epsilon =\pm 1.$ In the simplest case with $\Upsilon
_{4}=0$ the equation (\ref{ep1a}) is solved by arbitrary two functions $%
g_{2}(x^{3})$ and $g_{3}(x^{2}).$

\item For $\Upsilon _{2}=0,$ the equation (\ref{ep2a}) relates two functions
$h_{4}\left( x^{i},v\right) $ and $h_{5}\left( x^{i},v\right) $ following
two possibilities:

a) to compute
\begin{eqnarray}
\sqrt{|h_{5}|} &=&h_{5[1]}\left( x^{i}\right) +h_{5[2]}\left( x^{i}\right)
\int \sqrt{|h_{4}\left( x^{i},v\right) |}dv,~h_{4}^{\ast }\left(
x^{i},v\right) \neq 0;  \notag \\
&=&h_{5[1]}\left( x^{i}\right) +h_{5[2]}\left( x^{i}\right) v,\ h_{4}^{\ast
}\left( x^{i},v\right) =0,  \label{p2}
\end{eqnarray}%
for some functions $h_{5[1,2]}\left( x^{i}\right) $ stated by boundary
conditions;

b) or, inversely, to compute $h_{4}$ for a given $h_{5}\left( x^{i},v\right)
,h_{5}^{\ast }\neq 0,$%
\begin{equation}
\sqrt{|h_{4}|}=h_{[0]}\left( x^{i}\right) (\sqrt{|h_{5}\left( x^{i},v\right)
|})^{\ast },  \label{p1}
\end{equation}%
with $h_{[0]}\left( x^{i}\right) $ given by boundary conditions. We note
that the sourceless equation (\ref{ep2a}) is satisfied by arbitrary pairs of
coefficients $h_{4}\left( x^{i},v\right) $ and $h_{5[0]}\left( x^{i}\right)
. $ Solutions with $\Upsilon _{2}\neq 0$ can be generated by an ansatz of
type
\begin{equation}
h_{5}[\Upsilon _{2}]=h_{5},h_{4}[\Upsilon _{2}]=\varsigma _{4}\left(
x^{i},v\right) h_{4},  \label{auxf02}
\end{equation}%
where $h_{4}$ and $h_{5}$ are related by formula (\ref{p2}), or (\ref{p1}).
Substituting (\ref{auxf02}), we obtain%
\begin{equation}
\varsigma _{4}\left( x^{i},v\right) =\varsigma _{4[0]}\left( x^{i}\right)
-\int \Upsilon _{2}(x^{2},x^{3},v)\frac{h_{4}h_{5}}{4h_{5}^{\ast }}dv,
\label{auxf02a}
\end{equation}%
where $\varsigma _{4[0]}\left( x^{i}\right) $ are arbitrary functions. We
have to put $\varsigma _{4[0]}\left( x^{i}\right) =1$ in order to have
compatibility with the sourceless case $\Upsilon _{2}\rightarrow 0.$

\item The exact solutions of (\ref{ep3a}) for $\beta \neq 0$ are defined
from an algebraic equation, $w_{i}\beta +\alpha _{i}=0,$ where the
coefficients $\beta $ and $\alpha _{i}$ are computed as in formulas (\ref%
{abc}) by using the solutions for (\ref{ep1a}) and (\ref{ep2a}). The general
solution is
\begin{equation}
w_{k}=\partial _{k}\ln [\sqrt{|h_{4}h_{5}|}/|h_{5}^{\ast }|]/\partial
_{v}\ln [\sqrt{|h_{4}h_{5}|}/|h_{5}^{\ast }|],  \label{w}
\end{equation}%
with $\partial _{v}=\partial /\partial v$ and $h_{5}^{\ast }\neq 0.$ If $%
h_{5}^{\ast }=0,$ or even $h_{5}^{\ast }\neq 0$ but $\beta =0,$ the
coefficients $w_{k}$ could be arbitrary functions on $\left( x^{i},v\right)
. $ \ For the vacuum Einstein equations this is a degenerated case imposing
the the compatibility conditions $\beta =\alpha _{i}=0,$ which are
satisfied, for instance, if the $h_{4}$ and $h_{5}$ are related as in the
formula (\ref{p1}) but with $h_{[0]}\left( x^{i}\right) =const.$

\item Having defined $h_{4}$ and $h_{5}$ and computed $\gamma $ from (\ref%
{abc}) we can solve the equation (\ref{ep4a}) by integrating on variable ''$%
v $'' the equation $n_{i}^{\ast \ast }+\gamma n_{i}^{\ast }=0.$ The exact
solution is
\begin{eqnarray}
n_{k} &=&n_{k[1]}\left( x^{i}\right) +n_{k[2]}\left( x^{i}\right) \int
[h_{4}/(\sqrt{|h_{5}|})^{3}]dv,~h_{5}^{\ast }\neq 0;  \notag \\
&=&n_{k[1]}\left( x^{i}\right) +n_{k[2]}\left( x^{i}\right) \int
h_{4}dv,\qquad ~h_{5}^{\ast }=0;  \label{n} \\
&=&n_{k[1]}\left( x^{i}\right) +n_{k[2]}\left( x^{i}\right) \int [1/(\sqrt{%
|h_{5}|})^{3}]dv,~h_{4}^{\ast }=0,  \notag
\end{eqnarray}%
for some functions $n_{k[1,2]}\left( x^{i}\right) $ stated by boundary
conditions.

\item The exact solution of (\ref{confeq}) is given by some arbitrary
functions $\zeta _{i}=\zeta _{i}\left( x^{i},v\right) $ if \ both $\partial
_{i}\omega =0$ and $\omega ^{\ast }=0,$ we chose $\zeta _{i}=0$ for $\omega
=const,$ and
\begin{eqnarray}
\zeta _{i} &=&-w_{i}+(\omega ^{\ast })^{-1}\partial _{i}\omega ,\quad \omega
^{\ast }\neq 0,  \label{confsol} \\
&=&(\omega ^{\ast })^{-1}\partial _{i}\omega ,\quad \omega ^{\ast }\neq 0,%
\mbox{ for vacuum solutions}.  \notag
\end{eqnarray}%
\bigskip
\end{enumerate}

\section{Solitonic deformations with nontrivial conformal factors}

\label{ap2}We illustrate here that similar methods, applied in Section \ref%
{sexsol} can be used for generating d--metrics of type (\ref{ansatzfd}) with
nontrivial conformal factors $\omega .$

\subsection{Almost stationary metrics}

Let we consider how the Schwarzschild metric (\ref{schw5}) can deformed
nonholonomically to a d--metric, (\ref{ansatzfd})) with nontrivial conformal
factor $\omega ^{2}$ for a set of local coordinates $(x^{\alpha
}=(t,p,\theta ),$ $u^{a}=(v=\varphi ,$ $\chi ).$ We consider deformations of
type%
\begin{eqnarray}
q_{1} &=&1\rightarrow g_{1}=1,\ q_{2}=-\frac{a^{2}(p)}{b^{2}(p)}\rightarrow
g_{2}=q_{2}(p)\ \eta _{2}(p,\theta ),\   \notag \\
q_{3} &=&-\frac{a^{2}(p)}{b^{2}(p)}\rightarrow g_{3}=q_{3}(p)\ \eta
_{3}(p,\theta ),  \label{coefsol3} \\
q_{4} &=&-\frac{a^{2}(p)}{b^{2}(p)}\sin ^{2}\theta \rightarrow
h_{4}=q_{4}(p,\theta )\eta _{4}(p,\theta ,\varphi ),  \notag \\
\ q_{5} &=&\epsilon b^{-2}(p)\rightarrow h_{5}=q_{5}(p)\eta _{5}(p,\theta
,\varphi ),  \notag
\end{eqnarray}%
with nontrivial deformations of the conformal factor
\begin{equation*}
\omega _{0}^{2}=b^{2}(p)\rightarrow \omega ^{2}=\eta _{0}(p,\theta ,\varphi
)\omega _{0}^{2}(p)
\end{equation*}%
where the non--deformed values are stated by the coefficients (\ref%
{schw5coef}). The ''polarization'' functions $\eta _{2,3}(p,\theta )$ have
to be found as a solution of type (\ref{auxeq01}), (\ref{auxeq01a}), or (\ref%
{auxeq01b}), depending explicitly of the type of source $\Upsilon
_{4}(p,\theta )$ and vacuum boundary conditions. The ''polarization''
functions $\eta _{4,5}(p,\theta ,\varphi )$ are solutions of the equations (%
\ref{p1}), or (\ref{p2}), and (\ref{auxf02}), in their turn depending on the
type of source $\Upsilon _{2}(p,\theta ,\varphi )$ and vacuum boundary
conditions. This class of solutions can be represented in the form (\ref%
{gensol1})
\begin{eqnarray}
\delta s^{2} &=&\eta _{0}(p,\theta ,\varphi )\omega _{0}^{2}(p)[(dt)^{2}-%
\frac{a^{2}(p)}{b^{2}(p)}\eta _{2}(p,\theta )(dp)^{2}-\frac{a^{2}(p)}{%
b^{2}(p)}\eta _{3}(p,\theta )(d\theta )^{2}-  \notag \\
&&h_{0}^{2}(p,\theta )\left[ f^{\ast }\left( p,\theta ,\varphi \right) %
\right] ^{2}|\varsigma _{\Upsilon }\left( p,\theta ,\varphi \right) |\left(
\delta \varphi \right) ^{2}+\epsilon f^{2}\left( p,\theta ,\varphi \right)
\left( \delta \chi \right) ^{2}],  \label{sol3}
\end{eqnarray}%
for
\begin{eqnarray*}
\delta \varphi &=&d\varphi +w_{1}\left( t,p,\theta ,\varphi \right)
dt+w_{2}\left( t,p,\theta ,\varphi \right) dp+w_{3}\left( t,p,\theta
,\varphi \right) d\theta ,\  \\
\delta \chi &=&d\chi +n_{1}\left( t,p,\theta ,\varphi \right) dt+n_{2}\left(
t,p,\theta ,\varphi \right) dp+n_{3}\left( t,p,\theta ,\varphi \right)
d\theta ,
\end{eqnarray*}%
where we parametrize
\begin{equation*}
f^{2}\left( p,\theta ,\varphi \right) =b^{-2}(p)\eta _{5}(p,\theta ,\varphi
),h_{0}^{2}(p,\theta )=a^{2}(p))
\end{equation*}%
and
\begin{equation*}
h_{0}^{2}(p,\theta )\left[ f^{\ast }\left( p,\theta ,\varphi \right) \right]
^{2}|\varsigma _{\Upsilon }\left( p,\theta ,\varphi \right) |=\frac{a^{2}(p)%
}{b^{2}(p)}\sin ^{2}\theta \ \eta _{4}(p,\theta ,\varphi )
\end{equation*}%
with the N--connection coefficients $w_{k}$ and $n_{k}$ computed,
respectively, as certain integrals (\ref{gensol1w}) and (\ref{gensol1n}). We
call such spacetimes to be almost stationary because the d--metric
coefficients do not depend on time coordinate $t$ but the N--connection
coefficients and related nonholonomic frames of reference may posses a such
dependence. For a stated solitonic form for $h_{5}=h_{5}^{stn}=b^{-2}(p)\eta
_{5}^{stn},$ with $b^{2}(p)$ taken as for the Schwarzschild metric, $h_{4}$
can be computed
\begin{equation*}
h_{4}=h_{4}^{stn}=h_{[0]}^{2}\left[ \left( \sqrt{|h_{5}^{stn}(p,\theta
,\varphi )|}\right) ^{\ast }\right] ^{2}
\end{equation*}%
where $h_{[0]}$ is a constant (see formula (\ref{p1}) in the Appendix).

We satisfy the conditions (\ref{conf1}) if we choose any conformal factor
\begin{equation*}
\omega =\eta _{0}(p,\theta ,\varphi )\omega _{0}^{2}(p)=\left(
h_{4}~^{stn}\right) ^{p_{2}/p_{1}}
\end{equation*}%
for some integers $p_{1}$ and $p_{2}$ and $\ $defining $\zeta _{i}$ as
solutions of the equations \
\begin{equation*}
\partial _{i}\omega -(w_{i}+\zeta _{i})\omega ^{\ast }=0
\end{equation*}
for given solitonic values $w_{i}^{stn}$ and $\omega .$

The nontrivial anchor coefficients are defined by the values $q^{\alpha
}=1/q_{\alpha }$ and $\eta _{\beta }=1/\eta ^{\beta },$ stated by (\ref%
{coefsol3}), introduced in formulas (\ref{anch5}) for $\left( \ \widehat{%
\mathbf{\rho }}_{a}^{i}\right) ^{2}$ corresponding to $H_{a}^{i}(p,\theta
,\varphi ).$

\subsection{Metrics with explicit extra dimension polarization}

This type of deformations transforms the metric (\ref{schw5}) into a
d--metric (\ref{ansatzfd}) via re--parametrizations%
\begin{eqnarray}
q_{1} &=&1\rightarrow g_{1}=1,\ q_{2}=-\frac{a^{2}(p)}{b^{2}(p)}\rightarrow
g_{2}=q_{2}(p)\ \eta _{2}(p,\theta ),\   \notag \\
q_{3} &=&-\frac{a^{2}(p)}{b^{2}(p)}\rightarrow g_{3}=q_{3}(p)\ \eta
_{3}(p,\theta ),  \label{coefsol4} \\
\ q_{4} &=&\epsilon b^{-2}(p)\rightarrow h_{4}=q_{4}(p)\eta _{4}(p,\theta
,\chi ),  \notag \\
q_{5} &=&-\frac{a^{2}(p)}{b^{2}(p)}\sin ^{2}\theta \rightarrow
h_{5}=q_{5}(p,\theta )\eta _{5}(p,\theta ,\chi ),  \notag
\end{eqnarray}%
for the local coordinates $x^{\alpha }=(t,p,\theta ),$ $u^{a}=(v=\chi ,$ $%
\varphi ).$Such data are stated for the ''anisotropic'' dependence on the
extra dimension coordinate $\chi $ with the ''polarization'' functions $\eta
_{4,5}(p,\theta ,t)$ being solutions of the equations (\ref{p1}), or (\ref%
{p2}), and (\ref{auxf02}). This class of d--metrics (\ref{gensol1}) is
written%
\begin{eqnarray*}
\delta s^{2} &=&\eta _{0}(p,\theta ,\chi )\omega _{0}^{2}(p)[(dt)^{2}-\frac{%
a^{2}(p)}{b^{2}(p)}\eta _{2}(p,\theta )(dp)^{2}-\frac{a^{2}(p)}{b^{2}(p)}%
\eta _{3}(p,\theta )(d\theta )^{2}+ \\
&&\epsilon h_{0}^{2}(p,\theta )\left[ f^{\ast }\left( p,\theta ,\chi \right) %
\right] ^{2}|\varsigma _{\Upsilon }\left( p,\theta ,\chi \right) |\left(
\delta \chi \right) ^{2}-f^{2}\left( p,\theta ,\chi \right) \left( \delta
\varphi \right) ^{2}],
\end{eqnarray*}%
for
\begin{eqnarray*}
\delta \chi &=&d\chi +w_{1}\left( t,p,\theta ,\chi \right) dt+w_{2}\left(
t,p,\theta ,\chi \right) dp+w_{3}\left( t,p,\theta ,\chi \right) d\theta , \\
\delta \varphi &=&d\varphi +n_{1}\left( t,p,\theta ,\chi \right)
dt+n_{2}\left( t,p,\theta ,\chi \right) dp+n_{3}\left( t,p,\theta ,\chi
\right) d\theta ,
\end{eqnarray*}%
where we parametrize
\begin{equation*}
f^{2}\left( p,\theta ,\chi \right) =\frac{b^{2}(p)}{a^{2}(p)}\sin ^{2}\theta
\ \eta _{5}(p,\theta ,\chi ),h_{0}^{2}(p,\theta )=b^{-2}(p)
\end{equation*}%
and
\begin{equation*}
h_{0}^{2}(p,\theta )\left[ f^{\ast }\left( p,\theta ,\chi \right) \right]
^{2}|\varsigma _{\Upsilon }\left( p,\theta ,\chi \right) |=b^{-2}(p)\ \eta
_{4}(p,\theta ,\chi )
\end{equation*}%
with the N--connection coefficients $w_{k}$ and $n_{k}$ computed
respectively by the integrals (\ref{gensol1w}) and (\ref{gensol1n}).

The solitonic background of the d--metric (\ref{sol2}) is given by $%
h_{5}=h_{5}^{stn}=-\left[ b^{2}(p)\sin ^{2}\theta /a^{2}(p)\right] \ \eta
_{5}^{stn}$ and
\begin{equation*}
h_{4}=h_{4}^{stn}=h_{[0]}^{2}\left[ \left( \sqrt{|h_{5}^{stn}(p,\theta ,\chi
)|}\right) ^{\ast }\right] ^{2}
\end{equation*}%
defined by the 3D solitonic equation (\ref{kdp}), or (\ref{sineq}), for the
new set of coordinates, when $h_{[0]}=const.$ This allows to define $\eta
_{4}^{stn}$ $(p,\theta ,\chi )$ and $f^{stn}(p,\theta ,\chi ),$ which (by
using the $f$--function) result in off--diagonal terms (i. e. in
N--connection coefficients) $w_{k}^{stn}\left( t,p,\theta ,\chi \right) $ (%
\ref{gensol1w}) and $n_{k}^{stn}\left( t,p,\theta ,\chi \right) $ (\ref%
{gensol1n}).

We satisfy the conditions (\ref{conf1}) if we choose a conformal factor
\begin{equation*}
\omega ^{2}=\eta _{0}(p,\theta ,\chi )\omega _{0}^{2}(p)=\left(
h_{4}~^{stn}\right) ^{p_{2}/p_{1}}
\end{equation*}%
for some integers $p_{1}$ and $p_{2}$ and $\ $defining $\zeta _{i}$ as
solutions of the equations \
\begin{equation*}
\partial _{i}\omega -(w_{i}+\zeta _{i})\omega ^{\ast }=0
\end{equation*}%
for given solitonic values $w_{i}=w_{i}^{stn}$ and $\omega =\omega ^{stn}.$

The nontrivial anchor coefficients are defined by the values $q^{\alpha
}=1/q_{\alpha }$ and $\eta _{\beta }=1/\eta ^{\beta },$ stated by (\ref%
{coefsol4}), introduced in formulas (\ref{anch5}) for $\left( \ \widehat{%
\mathbf{\rho }}_{a}^{i}\right) ^{2}$ with $H_{a}^{i}(p,\theta ,\chi ).$

\end{document}